\documentclass[superscriptaddress,twocolumn,notitlepage,prx,longbibliography]{revtex4-2}
\usepackage[colorlinks=true,citecolor=blue,urlcolor=black]{hyperref}
\usepackage{graphicx}
\usepackage{dcolumn}
\usepackage{bm}
\usepackage{amsthm}
\usepackage{amsmath}
\usepackage{amssymb}
\usepackage{color}
\usepackage{multirow}
\usepackage{algorithm}
\usepackage{algorithmicx}
\usepackage{algpseudocode}
\usepackage{subfigure}
\usepackage{hyperref}
\usepackage{wasysym}
\usepackage{xcolor}
\usepackage{overpic}
\usepackage{color}
\usepackage{lineno}

\newtheorem{theorem}{Theorem}
\newtheorem{remark}{Remark}
\newtheorem{corollary}{Corollary}

\definecolor{Red}{rgb}{1,0,0}
\def\vec#1{{\bm #1}}

\def\ket#1{| #1 \rangle}
\def\bra#1{\langle #1 |}

\def\RR{\mathbb{R}}

\def\Tr{\operatorname{Tr}}

\def\R{\mathbb{R}}

\def\RR{\mathbb{R}}
\def\CC{\mathbb{C}}

\begin{document}


\title{A duplication-free quantum neural network for universal approximation}

\author{Xiaokai Hou}
\affiliation{Institute of Fundamental and Frontier Sciences, University of Electronic Science and Technology of China, Chengdu, 610051, China}

\author{Guanyu Zhou}
\email{zhoug@uestc.edu.cn}
\affiliation{Institute of Fundamental and Frontier Sciences, University of Electronic Science and Technology of China, Chengdu, 610051, China}

\author{Qingyu Li}
\affiliation{Institute of Fundamental and Frontier Sciences, University of Electronic Science and Technology of China, Chengdu, 610051, China}

\author{Shan Jin}
\affiliation{Institute of Fundamental and Frontier Sciences, University of Electronic Science and Technology of China, Chengdu, 610051, China}

\author{Xiaoting Wang}
\email{xiaoting@uestc.edu.cn}
\affiliation{Institute of Fundamental and Frontier Sciences, University of Electronic Science and Technology of China, Chengdu, 610051, China}

\begin{abstract}
The universality of a quantum neural network refers to its ability to approximate arbitrary functions and is a theoretical guarantee for its effectiveness. A non-universal neural network could fail in completing the machine learning task. One proposal for universality is to encode the quantum data into identical copies of a tensor product, but this will substantially increase the system size and the circuit complexity. To address this problem, we propose a simple design of a duplication-free quantum neural network whose universality can be rigorously proved. Compared with other established proposals, our model requires significantly fewer qubits and a shallower circuit, substantially lowering the resource overhead for implementation. It is also more robust against noise and easier to implement on a near-term device. Simulations show that our model can solve a broad range of classical and quantum learning problems, demonstrating its broad application potential. 
\end{abstract}

\maketitle

\section{Introduction}

Machine learning (ML) is a powerful data-analyzing tool that has generated a series of impactful results in image recognition~\cite{8237819,Shi2017AnET,8308186}, natural language processing~\cite{goldberg2016primer,goldberg2017neural,collobert2008unified}, self-driving~\cite{nugraha2017towards,do2018real,bojarski2017explaining}, etc. Meanwhile, quantum machine learning has become an emerging interdisciplinary subject combining machine learning with quantum computing. It studies two fundamental questions~\cite{biamonte2017quantum}, one on applications of classical ML to quantum problems~\cite{doi:10.1126/science,PhysRevX.7.021021,garrido2021augmenting,schutt2019unifying,gao2017efficient,paruzzo2018chemical}, and the other on implementations of ML algorithms on a quantum processor. Recent works point out that implementing ML algorithms on a quantum computer to process quantum data requires exponentially fewer experiments than dealing with the same task on a classical computer~\cite{doi:10.1126/science.abn7293}. In addition, it has been shown that many ML algorithms can be implemented on a quantum machine with certain levels of advantage, ranging from data fitting~\cite{PhysRevLett.109.050505}, support vector machine~\cite{rebentrost2014quantum}, nearest neighbors~\cite{10.5555/2871393.2871400}, principle component analysis~\cite{lloyd2014quantum}, and linear regression~\cite{PhysRevA.94.022342}, to PageRank algorithm~\cite{paparo2012google}, reinforcement learning~\cite{4579244,saggio2021experimental,PhysRevX.4.031002}, anomaly detection~\cite{PhysRevA.97.042315} and Boltzmann machine~\cite{PhysRevX.8.021050}. For other ML topics, especially the quantum neural networks (QNNs), the existence of quantum advantage is largely unknown~\cite{aaronson2015read}, although an interesting result on the prediction advantage for QNNs has been recently achieved~\cite{Huang_2021}. 

A neural network (NN) is a parameterized composite mapping comprised of activation functions and is very powerful in data fitting. A QNN is an NN implemented on a quantum computer~\cite{schuld2014quest} and there are various proposals to design it, such as the variational QNN~\cite{mitarai2018quantum,schuld2020circuit}. Besides quantum advantage, designing an effective and efficient QNN is also important. Not every QNN design effectively solves the given ML problem, and the concept of \emph{universal approximation} is introduced to explain why some neural networks fail to work~\cite{hornik1989,LESHNO1993861}. Universal approximation, or the \emph{universality} of an NN, refers to its ability to approximate arbitrary functions. For classical NNs, universality is easy to achieve; for QNNs, a smart design is required to make them universal and practical. In fact, the universality of NNs is closely related to the nonlinear activation functions, and the key point of designing a universal QNN is to generate the desired nonlinearity for function fitting. One method is to construct the quantum neurons, which are the building blocks of some QNNs~\cite{cao2017quantum,yan2020nonlinear,tacchino2019artificial,kristensen2021artificial,Torrontegui_2019,beer2020training}; the other method is to construct a variational QNN from a parameterized quantum circuit, and the nonlinearity can be achieved by duplicating the quantum data into a tensor product of multiple copies~\cite{SCHULD2015660,wan2017quantum}. Examples of the latter method include the quantum circuit learning~(QCL) algorithm~\cite{mitarai2018quantum} and the circuit-centric quantum classifiers (CCQ) algorithm~\cite{schuld2020circuit}. So far, the universality of the QCL algorithm has been proved using the Weierstrass theorem, while the universality of the CCQ algorithm remains open. For both QCL and CCQ, approximating a highly-nonlinear function would require a tensor product of many data-encoding subsystems, resulting in a large overall system size with a considerable circuit complexity, conflicting with the principle of NISQ computing where a relatively small quantum system with a shallow circuit is preferred.

To address this problem, we propose a duplication-free quantum neural network (DQNN) based on a variational quantum circuit and rigorously prove its universality. Our model utilizes the classical sigmoid function to generate nonlinearity without duplicating the quantum data into a tensor product of multiple copies. Compared with the CCQ or QCL algorithms, our DQNN significantly reduces the system size and gate complexity, and hence the overall noise effect. Therefore, it is more likely to be implemented on near-term devices. Numerical simulations show that our DQNN outperforms the other two variational QNN algorithms with better performance on typical regression and classification problems and is more robust against noise. In addition, through solving a broad range of classical and quantum learning problems, our model has well demonstrated its wide application potential.

\section{Two types of designs for universal approximation}

Before we discuss the DQNN design, let's briefly review two common types of designs for the universal approximation of a classical NN. It turns out that universality is closely related to different kinds of convergence. Type I approximation relies on pointwise convergence; a typical example is the polynomial approximation, using the Weierstrass theorem and polynomials to approximate arbitrary functions; Type II approximation depends on the convergence in Lebesgue integration, e.g., $L^2$-norm convergence. For polynomial approximation, there are several limitations. First, the Weierstrass theorem is only valid on a compact set, while many NN problems are defined on unbounded set, on which polynomials could fail to approximate a Lebesgue integrable function. For example, the function $f(x)=e^{-x}$ defined on $[0,+\infty)$ can not be well approximated by polynomials since the latter diverges to $\pm\infty$ while $e^{-x}$ vanishes as $x \rightarrow \infty$. Second, polynomial approximation has a numerical problem called Runge's phenomenon~\cite{runge} when the polynomial order becomes very large, which is why using high-order polynomials for function fitting is not favored in practical use. Due to these limitations, polynomials are ineffective in designing universal NNs. It has been proved that the necessary and sufficient condition for a universal feedforward NN is that its activation functions are not polynomial~\cite{LESHNO1993861}. 
In comparison, Type II approximation utilizes basis functions such as the hat function in finite-element method~\cite{brenner_finite_element} or the sigmoid/ReLU function composite with linear function in neural networks~\cite{Goodfellow-et-al-2016}. Usually, the optimal approximation is the solution to a minimization problem, and the error is evaluated in a suitable function norm. In particular, universal NNs based on $L^2$-norm approximation are widely applied and found to be very effective. Such examples include the ones with activation functions, such as the sigmoid or the ReLU functions. Hence, this work aims to discuss how to construct a universal and effective QNN based on $L^2$ approximation, using the trick analogous to the sigmoid function instead of the polynomials. We expect that such QNN is effective in a broad range of applications compared to the QNN based on polynomial approximation. 

\section{DQNN and its universality}

\begin{figure*}[htp]
    \setlength{\belowcaptionskip}{-3mm}
    \centering
    \subfigure[large][]{\includegraphics[width=0.35\textwidth]{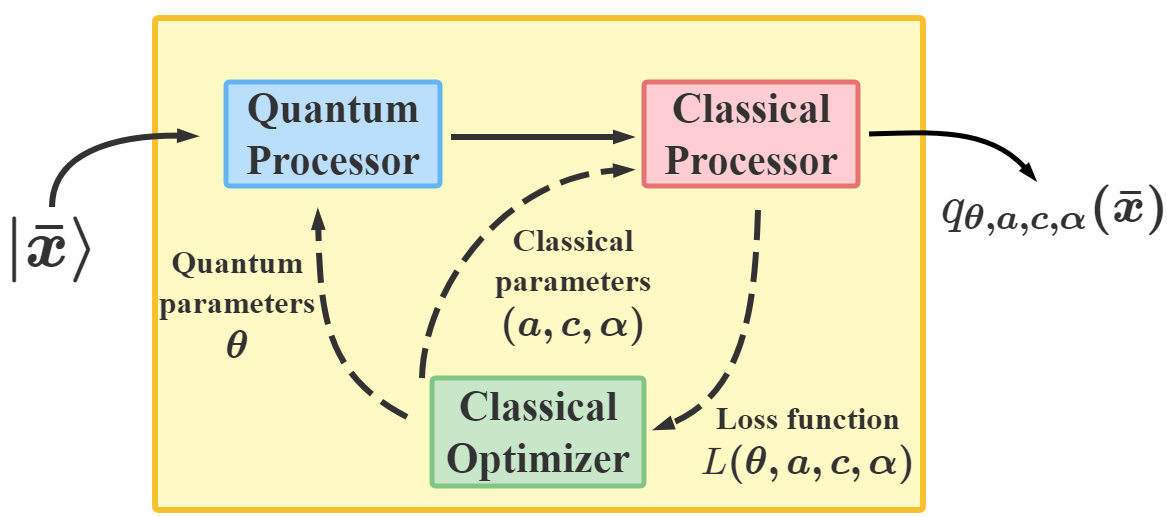}
    \label{frame}}
    \subfigure[]{\includegraphics[width=0.20\textwidth]{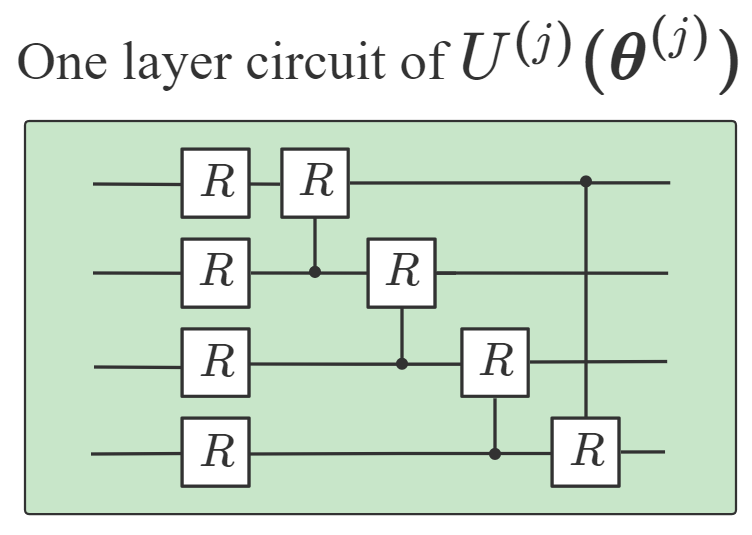}\label{ansatz}}
    \\
    \subfigure[]{\includegraphics[width=0.58\textwidth]{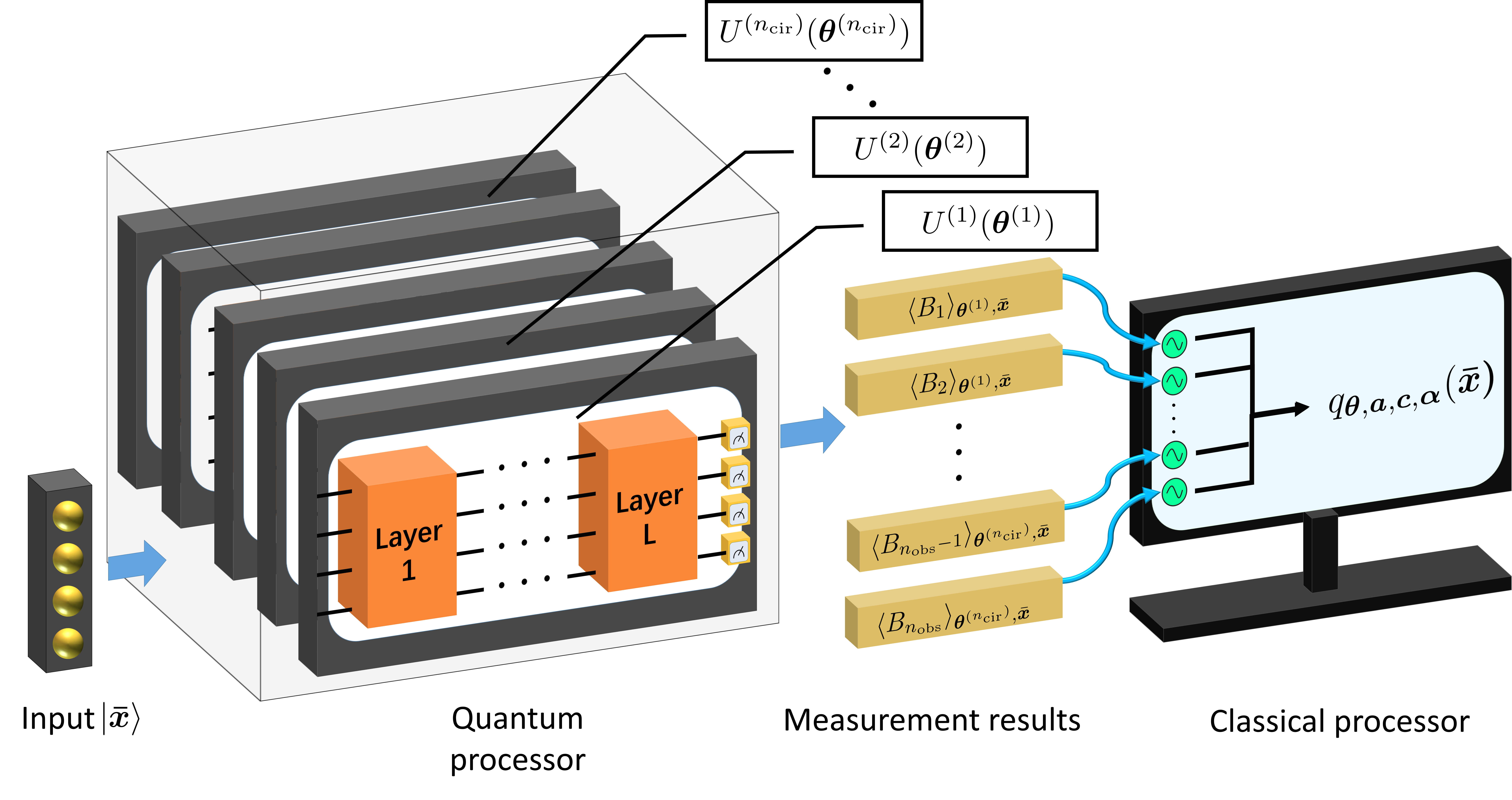}\label{circuit}}
    \caption{\textbf{(a)} The framework of the DQNN consists of three parts, a quantum processor, a classical processor and a classical optimizer. Measurement results of a quantum processor are the inputs of a classical processor. The parameters in the quantum and classical processors are updated by a classical optimizer. \textbf{(b)} One layer variational quantum circuit of $U^{(j)}(\bm{\theta}^{(j)})$. \textbf{(c)} The transformation from $|\bar{\vec{x}}\rangle$ to $q_{\bm{\theta},\bm{a},\bm{c},\bm{\alpha}}(\bar{\bm{x}})$. Yellow balls represent the quantum qubits; green circles represent the classical sigmoid function.}
    \label{Fig1}
\end{figure*}

Mathematically, a variational QNN is a mapping $q_{\vec\theta}: \vec x\in \RR^d \to q_{\vec\theta}(\bm{x})\in \RR$ built on a parameterized quantum circuit $U(\vec\theta)$. It involves a hybrid quantum-classical algorithm that utilizes a quantum processor to generate the output $q_{\vec\theta}(\vec x)$ for each input $\vec x$, and a classical processor to optimize $\vec\theta$ to solve learning problems. Specifically, given an unknown function $f: \vec x \to f(\bm{x})$, and a data set $D=\{(\bm{x}^{(i)},y^{(i)})\}^m_{i=1}$, where $y^{(i)}=f(\bm{x}^{(i)})$ is the label of $\vec{x}^{(i)}$, the aim of QNN is to find an optimal $\vec\theta^*$ such that $q_{\vec\theta^*}$ best approximates the target function $f$. Our DQNN model consists of three parts, a quantum processor, a classical processor and a classical optimizer, as illustrated in Fig.~\ref{frame}. First, each data point $\vec x=[x_1,x_2,\cdots,x_d]^T\in \RR^d$ is encoded into a state of an $n$-qubit register ($n= \lceil \log d \rceil$):

\begin{equation}
    |\bar{\vec{x}}\rangle=\frac{1}{\gamma}[x_1,x_2,\cdots,x_d,\tilde{x},0,\cdots0]^T \in \CC_2^{\otimes n},
\end{equation}
using the amplitude encoding method~\cite{PhysRevA.83.032302}, where $\tilde{x}\equiv\frac{\|\bm{x}\|}{1+\|\bm{x}\|}$ is a padding term chosen by the user and $\gamma$ is a normalization factor (Appendix~\ref{Appendix_amplitude_encoding}). After initial state preparation, we apply to $|\bar{\vec{x}}\rangle$ a series of variational quantum circuits $\{U^{(j)}(\bm{\theta}^{(j)})\}_{j=1}^{n_{\text{cir}}}$ according to $|\bar{\bm{x}}_f^{(j)}\rangle=U^{(j)}(\bm{\theta}^{(j)})|\bar{\bm{x}}\rangle$, followed by measurements of a set of observables $\{B_i\}_{i=1}^{n_{\text{obs}}}$, to obtain the outcomes $\langle B_i\rangle_{\vec\theta^{(j)},\vec{\bar{x}}}=\mathrm{Tr}(|\bar{\bm{x}}_f^{(j)}\rangle\langle \bar{\bm{x}}_f^{(j)}|B_i)$. The choices of $\{B_i\}$ are not unique and they can be chosen randomly from the generalized Pauli basis $\{P_l\}_{l=1}^{4^n}$ where $P_l=A_{1}^{(l)}\otimes A_{2}^{(l)}\otimes\cdots\otimes A_{n}^{(l)}$ with $A_{k}^{(l)}\in\{I,X,Y,Z\}$, and $\{X,Y,Z\}$ are the Pauli matrices. $U^{(j)}(\vec\theta^{(j)})$ comprises $L$-layers of quantum circuits. Each layer consists of $n$ parameterized $R$-rotations (with one on each qubit), and $n$ parameterized controlled-$R$ gates, as shown in Fig.~\ref{ansatz}, with
\begin{equation*}
    R=R(\theta_1,\theta_2,\theta_3)=\left(
    \begin{matrix}
    e^{i\theta_2}\cos\theta_1 & e^{i\theta_3}\sin\theta_1\\
    -e^{-i\theta_3}\sin\theta_1 & e^{-i\theta_2}\cos\theta_1
    \end{matrix}
    \right).
\end{equation*}
Notice that the order of the $2n$ gates in each layer is not unique and can be chosen randomly~\cite{schuld2020circuit}. Next, based on $\langle B_i\rangle_{\vec\theta^{(j)},\vec{\bar{x}}}$ and the sigmoid function $\sigma(x)=1/(1+e^{-x})$, the classical processor computes and obtains the output:
\begin{equation}\label{qx}
    q_{\vec\theta,\vec a,\vec c,\vec \alpha}(\bar{\bm{x}}) \equiv \sum_{j=1}^{n_{\text{cir}}} \sum_{i=1}^{n_{\text{obs}}}\alpha_i^{(j)}\sigma(a_i^{(j)}(\langle B_i\rangle_{\vec\theta^{(j)},\bar{\bm{x}}}-c_i^{(j)}))
\end{equation}
 with $a_i^{(j)}>0$, $c_i^{(j)}\in[0,1]$ where $\bm{\theta}$ and $(\vec a,\vec c,\vec \alpha)$ are parameters to be trained. The entire process from $|\bar{\vec{x}}\rangle$ to $q_{\bm{\theta},\bm{a},\bm{c},\bm{\alpha}}(\bar{\bm{x}})$ is summarized in Fig.~\ref{circuit}. Finally, one can find the optimal values of $(\vec\theta,\vec a,\vec c,\vec \alpha)$ to solve the given learning problem, through gradient-based optimization using algorithms such as SGD~\cite{10.1007/978-3-7908-2604-3_16}, ADAM~\cite{DBLP:journals/corr/KingmaB14} or BFGS~\cite{buckley1985algorithm}. The benefit of such construction is that DQNN can be proved to be universal based on $L^2$ approximation, which is summarized in the following theorem: 
\begin{theorem}\label{theorem1}
Let $\bar{G}$ be a subset of the complex sphere $\mathbb{S}$ in $\mathbb{C}_2^{\otimes n}$, and $f(\bar{\vec{x}}): \bar{G} \rightarrow \mathbb{R}$ be an arbitrary square-integrable function on $\bar{G}$. We define the following parameterized functions: 
\begin{align}\label{sigmoid_fn}
   q_{\bm{z}_1,\cdots,\bm{z}_{n_s},\bm{a},\bm{c},\bm{\alpha}}(\bar{\bm{x}}) \equiv \sum_{j=1}^{n_s}\alpha_j\sigma(a_j(|\langle\bar{\bm{x}}|\bm{z}_j\rangle|^2-c_j))
\end{align}
and denote $Q(\bar{G})\equiv \{q_{\bm{z}_1,\cdots,\bm{z}_{n_s},\bm{a},\bm{c},\bm{\alpha}} \}$ as the set of all such functions, where 
$n_s\in\mathbb{N}$,  $\{\bm{z}_j \}_{j=1}^{n_s} \subset \mathbb{S}$, $\bm{a} \in\mathbb{R}_+^{ n_s}$, $\bm{c} \in [0,1]^{\otimes n_s}$, and $\bm{\alpha} \in \mathbb{R}^{ n_s}$. Then $Q(\bar{G})$ is dense in $L^2(\bar{G})$ in the following sense: for any $\epsilon>0$,
    \begin{equation}
        \int_{\bar{G}} | q_{\bm{z}_1,\cdots,\bm{z}_{n_s},\bm{a},\bm{c},\bm{\alpha}}(\bar{\bm{x}})-f(\bar{\bm{x}})|^2 d\mu < \epsilon.
    \end{equation}
\end{theorem}
The detailed proof of Theorem~\ref{theorem1} can be found in Appendix~\ref{Appendix:Proof_theorem1}. It turns out that any $q_{\bm{z}_1,\cdots,\bm{z}_{n_s},\bm{a},\bm{c},\bm{\alpha}}\in Q(\bar{G})$ in Eq.~(\ref{sigmoid_fn}) can be generated as the output of the DQNN in Eq.~(\ref{qx}). Specifically, given the parameters $(\bm{z}_1,\cdots,\bm{z}_{n_s},\bm{a},\bm{c},\bm{\alpha})$, we can design a DQNN with a single observable $B\equiv \ket{b}\bra{b}$, $n_{\text{obs}}=1$, and a series of quantum circuits $\{U^{(j)}(\bm{\theta}^{(j)})\}_{j=1}^{n_s}$ such that $U^{(j)}(\bm{\theta}^{(j)})\ket{b}=\ket{z_j}$. Then the output of the DQNN in Eq.~(\ref{qx}) is reduced to: 
\begin{align}
    q_{\vec\theta,\vec a,\vec c,\vec \alpha}(\bar{\bm{x}}) &= \sum_{j=1}^{n_s} \alpha_j\sigma(a_j(\langle B\rangle_{\vec\theta^{(j)},\bar{\bm{x}}}-c_j))  \nonumber\\
    &=\sum_{j=1}^{n_s}\alpha_j\sigma(a_j(|\langle\bar{\bm{x}}|\bm{z}_j\rangle|^2-c_j)) \nonumber\\
    &=q_{\bm{z}_1,\cdots,\bm{z}_{n_s},\bm{a},\bm{c},\bm{\alpha}}(\bar{\bm{x}}).
\end{align}

Thus we have proved the following corollary:
\begin{corollary}\label{corollary1}
The DQNN designed in Fig.~\ref{Fig1} (c) with the output in Eq.~(\ref{qx}) is universal. 
\end{corollary}
Notice that choosing multiple $\{U^{(j)}(\bm{\theta}^{(j)})\}_{j=1}^{n_{\text{cir}}}$ and a single observable $B$ is only a sufficient condition to prove the universality in Corollary~\ref{corollary1}. In practice, we often design a DQNN with a single $U(\bm{\theta})$ ($n_{\text{cir}}=1$) and a set of $\{B_i\}$ as the first trial to solve the given problem. If it is not good enough, we then construct a new DQNN with a larger $n_{\text{cir}}$. It turns out a DQNN with $n_{\text{cir}}=1$ is often sufficient for most learning problems we have studied. In addition, in the proof of Corollary~\ref{corollary1}, we have assumed that the variational circuits $\{U^{(j)}(\bm{\theta}^{(j)})\}$ with $U^{(j)}(\bm{\theta}^{(j)})\ket{b}=\ket{z_j}$ can be generated by our designed multi-layer variational circuits as shown in Fig.~\ref{Fig1} (b). More layers in $U^{(j)}(\bm{\theta}^{(j)})$ will make it more powerful to approximate arbitrary unitary gate, but will also make the optimization process computationally more expensive. Hence, in practice, we often construct a DQNN starting from a single $U(\bm{\theta})$ consisting of one or two layers. In the following, we will study the performance of DQNN with different values of $n_{\text{cir}}$, and for simplicity, we will denote the corresponding network as DQNN$_{n_{\text{cir}}}$.

\section{Performance comparison among QNN models}

In Section II, we have mentioned that universality serves as a necessary condition to build an effective QNN, and among the two types of universal design, an NN with $L^2$ convergence using sigmoid basis functions works more effectively in a broader range of applications. As shown above, the nonlinearity and the universality of DQNN originates essentially from the $L^2$ convergence of sigmoids functions, which is distinctive from other polynomial-approximation-based QNN models, including the CCQ~\cite{schuld2020circuit} and the QCL~\cite{mitarai2018quantum}, whose nonlinearity is mainly generated by duplicating the same data-encoding state into a tensor product of multiple subsystems. In this way, DQNN has significantly reduced the number of qubits required to construct a universal QNN. In the following, we will demonstrate the advantages of DQNN by comparing its performance with those of the two well-established QNN models, the CCQ and the QCL.

For an $n_{\text{tot}}$-qubit QNN model, we define its complexity as $C\equiv n_{\text{gate}}n_{\text{m}}$, where $n_\text{gate}$ denotes the number of quantum gates and $n_{\text{m}}$ denotes the number of quantum measurements. Notice that given a measurement precision, $n_{\text{m}}$ is proportional to the number of measurement observables $n_{\text{obs}}$. Hence, with a further assumption that the QNN variational circuit can be efficiently generated, i.e., $n_{\text{gate}}=O(\text{poly}\big(n_{\text{tot}})$, the overall complexity of the QNN then becomes $C= O(\text{poly}\big(n_{\text{tot}})n_{\text{obs}}\big)$. We hope to figure out the numbers of qubits and observables required to approximate an $M$-order polynomial function $f(\bm{x})=\sum_{i=0}^M c_i\bm{x}^i$ with $||\bm{x}||=1$ using CCQ, QCL and DQNN respectively. We denote by $n_{\text{data}}$ the number of qubits in the data-encoding register, and by $n_{\text{copy}}$ the number of copies of the data-encoding register used in a QNN model. To approximate such $f$, CCQ requires $n_{\text{copy}}=\frac{M}{2}$ duplicates of the data-encoding register $|\bar{\bm{x}}\rangle=\frac{1}{\gamma}\sum_{i=1}^{d+1} x_i|i\rangle$ such that the amplitude of the input state, $|\psi_{\vec{x}}\rangle=|\bar{\bm{x}}\rangle^{\otimes \frac{M}{2}}$, is an $\frac{M}{2}$-order polynomial of $\bm{x}$. After applying a circuit $U(\bm{\theta})$, we measure the expectation values of a given observable $B$ according to $\langle B\rangle_{\bm{\theta},\bm{x}}= \langle\psi_\vec{x}|U(\bm{\theta})^{\dagger} B U(\bm{\theta})|\psi_{\vec{x}}\rangle$. The output of the CCQ is an $M$-order polynomial of $\bm{x}$. Then the goal of CCQ is to find the best value $\bm{\theta}=\bm{\theta}^*$ through optimization such that $\bm{x}\rightarrow \langle B\rangle_{\bm{\theta}^*,\bm{x}}$ is the optimal function to approximate $\bm{x}\rightarrow f(\bm{x})$. The number of qubits in the data-encoding register is $n_{\text{data}}=\lceil \log d \rceil$ and the number of observables is $n_{\text{obs}}=1$. Hence the total complexity of CCQ is equal to $C_{\text{CCQ}}=O(\text{poly}(M\lceil \log d \rceil))$. For QCL, a $d$-dimensional classical data $\bm{x}$ is encoded into a $d$-qubit data-encoding register as $\rho(\boldsymbol{x})=\frac{1}{2^d}\otimes_{i=1}^d (\text{I}+x_i X+\sqrt{1-x_i^2} Z])$ using several single-qubit rotation gates. In order to generate an $M$-order polynomial, $\rho(\boldsymbol{x})$ is duplicated into $M$ identical copies in a tensor product as $\rho_{\text{in}}(\boldsymbol{x})=\otimes_{j=1}^M \rho(\boldsymbol{x})$. Applying a circuit $U(\bm{\theta})$ to $\rho_{\text{in}}$, we measure the expectation value of a given observable $B$. The outcome $\langle B \rangle_{\bm{\theta},\bm{x}}$ is an $M$-order polynomial of $\bm{x}$. Through  optimization, we can find the best value $\bm{\theta}^*=\bm{\theta}$ such that $\bm{x}\rightarrow \langle B\rangle_{\bm{\theta}^*,\bm{x}}$ will well approximate $\bm{x}\rightarrow f(\bm{x})$. The total number of qubits in QCL is $n_{\text{tot}}=Md$ and the number of observables is $n_{\text{obs}}=1$. Hence the total complexity of QCL is $C_{\text{QCL}}=O(\text{poly}(Md))$.

Different from the above two duplication-based QNN models, DQNN only uses $\lceil \log d \rceil$ qubits to encode $\bm{x}$ as $|\bar{\bm{x}}\rangle=\frac{1}{||\bm{x}||}\sum_{i=1}^{d+1} x_i|i\rangle$. We sequentially apply a set of variational quantum circuits $\{U^{(j)}(\bm{\theta}^{(j)})\}_{j=1}^{n_{\text{cir}}}$ to $|\bar{\bm{x}}\rangle$ and measure the expectation $\langle B_i\rangle_{\bm{\theta}^{(j)},\bar{\bm{x}}}=\langle\bar{\bm{x}}|U^{(j)\dagger}(\bm{\theta}^{(j)})B_iU^{(j)}(\bm{\theta}^{(j)})|\bar{\bm{x}}\rangle$ for a set of chosen observables $\{B_i\}_{i=1}^{n_{\text{obs}}}$. Next we apply the parameterized classical sigmoid function $\sigma$ and derive the final output $q_{\bm{\theta},\bm{a},\bm{c},\bm{\alpha}}$. Through the optimization, we can find the optimal values $(\bm{\theta}^*,\bm{a}^*,\bm{c}^*,\bm{\alpha}^*)$ to approximate the function $f$. The number of measurement observables $n_{\text{obs}}$ is determined by $f$. Hence the total complexity of DQNN is $C_{\text{DQNN}}=O(\text{poly}(\lceil \log d \rceil)n_{\text{obs}})$. The comparison is summarized in Table~\ref{compare}. It can be seen that for large $d$ and large $M$, our DQNN can significantly reduce the total complexity to generate nonlinearity. More details can be found in Appendix~\ref{Appendix:CCQ_QCL}.

\begin{table}[htb]
    \centering
    \setlength{\belowcaptionskip}{-0.4cm}
    \begin{tabular}{c|c|c|c}
    \hline
    \hline
    Algorithm &  $n_{\text{copy}}$  & $n_{\text{data}}$ & Complexity\\
    \hline
    QCL & $M$ & $d$ & $O(\text{poly}(Md))$\\
    \hline
    CCQ & $\frac{M}{2}$ & $\lceil \log d \rceil$ & $O(\text{poly}(M\lceil \log d \rceil))$\\
    \hline
    DQNN & $1$ & $\lceil \log d \rceil$ & $O(\text{poly}(\lceil \log d \rceil)n_{\text{obs}})$\\
    \hline
    \hline
    \end{tabular}
    \caption{The number of required qubits and the circuit complexity among three proposals to approximate an $M$-order polynomial function $f$ of $\bm{x}$.}
    \label{compare}
\end{table}

\begin{figure*}[htb]
    \centering
    \setlength{\belowcaptionskip}{-0.5cm}
    \subfigure[]
    {
     \begin{minipage}{.3\linewidth}\label{fit_com_dataset}
     \centering
      \includegraphics[width=4cm,height=3cm]{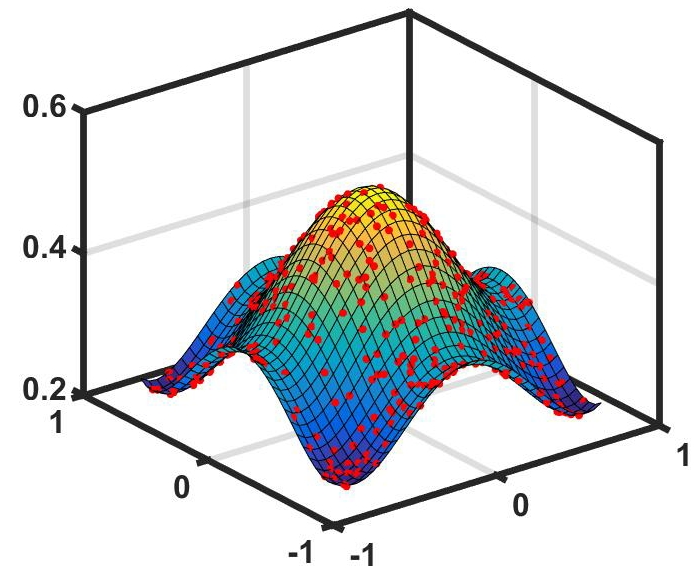}
     \end{minipage}
    }
    \subfigure[]
    {
     \begin{minipage}{.3\linewidth}\label{classification_com_dataset}
     \centering
      \includegraphics[width=3.2cm,height=3cm]{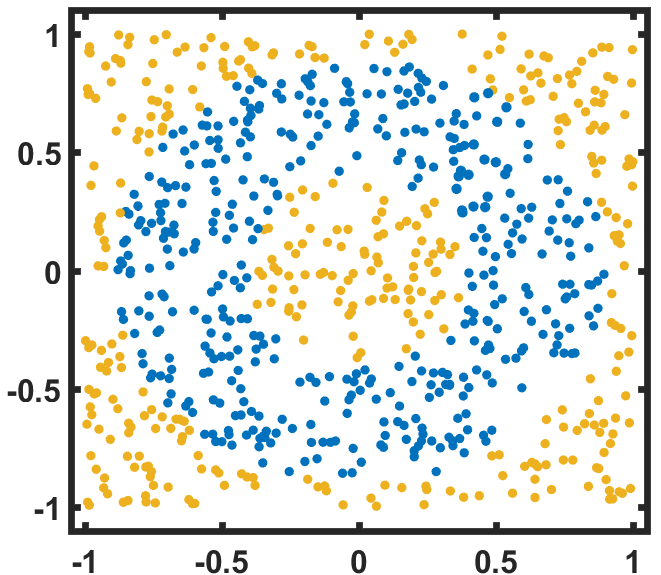}
     \end{minipage}
    }\\
    \subfigure[]{\includegraphics[width=0.6\textwidth]{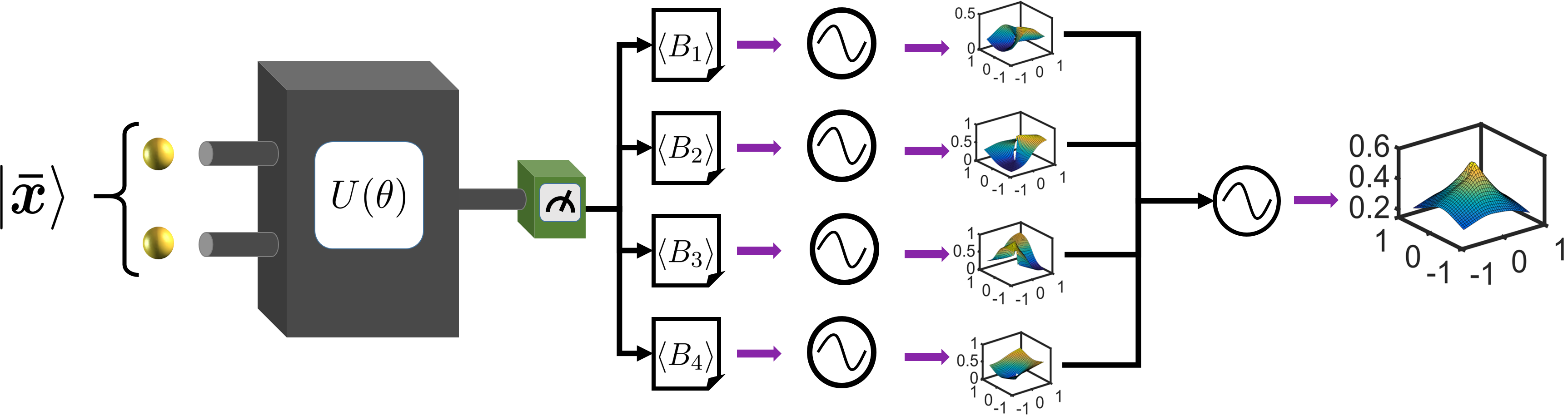}\label{DQNN_reg}}
   \caption{\textbf{(a)} The regression data set generated by a polynomial function. \textbf{(b)} The classification data set with the ring-shaped boundaries. The points in the yellow part are labeled $0$; the others are labeled $1$. \textbf{(c)} The approximation process of the DQNN$_1$ with one variational quantum circuit and $4$ observables in solving the regression problem based on the dataset in \textbf{(a)}. In the task, one more sigmoid function (the rightmost node) is added into the DQNN$_1$ to transform the output of DQNN$_1$ into a range $(0,1)$.}
\end{figure*}

To demonstrate that our DQNN has advantage in solving classical ML problems and in reducing circuit complexity, we apply DQNN, QCL, and CCQ to two typical supervised learning problems, a regression problem and a classification problem, and compare their performance. The first problem is learning from a data set $\{\vec{x}^{(i)},y^{(i)}\}$ to approximate a highly nonlinear polynomial $f(\vec{x})=(0.7156 - 1.0125 x_1^{2} + x_1^{4}) (0.7156 - 1.0125 x_2^{2} + x_2^{4})$ with $x_1,x_2\in[-0.8,0.8]^2$ (Fig.~\ref{fit_com_dataset}). For input $\vec{x}^{(i)}$, the output of the QNN is denoted as $q^{(i)}$. One can define the \textit{relative error} $\epsilon\equiv \frac{1}{N}\sum_i |\frac{q^{(i)}-y^{(i)}}{y^{(i)}}|$ to measure the performance of the QNN in solving regression problems. A tiny relative error implies that the QNN approximates the target function $f$ well. First, we run the hybrid optimization process for each model based on the data set. To show the effectiveness of the three QNN models, we choose different values of the number of layers for each model so that their circuit complexity $C$ is similar in size. We use the same definition of $n_{\text{copy}}$ as in the above section. Then, we compare the optimal relative errors achieved for each model. We choose two types of DQNN to solve the regression problem. One has a single-layer variational quantum circuit and multiple observables. We name it DQNN$_1$. The other has $4$ single-layer variational quantum circuits and one observable. We name it DQNN$_4$. We find that, with no duplicate, $n_{\text{copy}}=1$, the relative error achieved by DQNN is substantially lower than those achieved by QCL and CCQ, as shown in Table~\ref{table_regression}(a). If we add one more duplicate to both the QCL and the CCQ models, their optimal relative error will decrease, but their complexity will increase and surpass that of the DQNN. In addition, we implement such comparison for a few other examples, and simulation results suggest that DQNN can prominently reduce the circuit complexity to approximate highly nonlinear functions compared with QCL and CCQ. The approximation process of the DQNN$_1$ with $4$ observables is shown in Fig.~\ref{DQNN_reg}, and more details of the comparison can be found in Appendix~\ref{Appendix:regression}.

We apply the three models to a binary classification problem on a ring-shaped data set in the second task. The boundaries of the two sub-datasets are determined by six curves $x_1^2+x_2^2=0.16$, $x_1^2+x_2^2=0.81$, and $x_1=x_2=\pm 1$ (Fig.~\ref{classification_com_dataset}). One can use \textit{classification accuracy} to quantify the performance of QNN model in solving classification problems, which is defined as $\epsilon\equiv\frac{1}{N}\sum_i1_{\delta(q^{(i)}-y^{(i)})}$ where $q^{(i)}$ is the output of QNN, $y^{(i)}$ represents the label of $\vec{x}^{(i)}$ and $\delta$ is the Dirac delta function~\cite{bishop:2006:PRML}. Analogous to the first task, we choose the DQNN with one variational quantum circuit and multiple observables~(DQNN$_1$) and the DQNN with $4$ variational quantum circuits and one observable~(DQNN$_4$) for the classification task. We choose appropriate values for $n_{\text{copy}}$ and $n_{\text{lay}}$, so that the circuit complexity $C$ of the three models is similar in size. After optimizing the three QNN models, we find that accuracy achieved by the DQNN is substantially higher than those achieved by QCL and CCQ. In addition, increasing the number of duplicates does help QCL and CCQ to improve their optimal accuracy, but even a $5$-copy model (with 12 qubits) for QCL or CCQ cannot perform equally well as a DQNN with only two qubits (Table~\ref{table_regression}(b)). Next, we implement such a comparison for a few other examples. Simulation results suggest that DQNN can prominently reduce the circuit complexity compared to QCL and CCQ, but maintain a good performance in solving classification problems.

\begin{table*}[htp]
\centering
\subtable[Regression problem to approximate $f(\vec{x})=(0.7156 - 1.0125 x_1^{2} + x_1^{4}) (0.7156 - 1.0125 x_2^{2} + x_2^{4})$]{
\begin{tabular}{c|c|c|c|c|c|c|c}
\hline
\hline
 Model & $n_{\text{data}}$ & Copy number~($n_{\text{copy}}$) & \# Qubits~($n_{\text{tol}}$)  & \# Layers~($n_{\text{lay}}$) & \# Observables~($n_{\text{obs}}$)  & $C$ & Relative error \\
\hline
DQNN$_1$ & 2 & 1 & 2 & 1 & 4  & 48 & 6.79\%\\
\hline
DQNN$_4$ & 2 & 1 & 2 & 1 & 1  & 48 & 5.46\%\\
\hline
QCL & 2 & 1 & 2 & 4 & 1  & 48 & 8.21\%\\
\hline
CCQ & 2 & 1 & 2 & 4 & 1 & 51 & 12.92\%\\
\hline
QCL & 2 & 2 & 4 & 5 & 1 & 120 & 7.35\% \\
\hline
CCQ & 2 & 2 & 4 & 4 & 1 & 99 & 4.74\% \\
\hline
\hline
\end{tabular}
}
\subtable[A classification problem defined on a ring-shaped dataset]{
\begin{tabular}{c|c|c|c|c|c|c|c}
\hline
\hline
    Model & $n_{\text{data}}$ & Copy number~($n_{\text{copy}}$) & \# Qubits~($n_{\text{tol}}$)  & \# Layers~($n_{\text{lay}}$) & \# Observables~($n_{\text{obs}}$) & $C$ & Accuracy \\
    \hline
    DQNN$_1$ & 2 & 1 & 2 & 1 & 20  & 240 & 91.40\%\\
    \hline
    DQNN$_4$ & 2 & 1 & 2 & 1 & 1  & 48 & 97.63\%\\
    \hline
    QCL & 2 & 2 & 4 & 5 & 2 & 240 & 74.18\%\\
    \hline
    CCQ & 2 & 2 & 4 & 10 & 1 & 243 & 75.20\%\\
    \hline
    QCL & 2 & 6 & 12 & 10 & 2  & 1440 & 76.93\%\\
    \hline
    CCQ & 2 & 6 & 12 & 10 & 1 & 723 & 80.63\%\\
    \hline
    \hline
\end{tabular}
}
\qquad
\caption{A comparison of the performance among DQNN, QCL and CCQ in two classical ML problems. $n_{\text{data}}$ indicates the number of qubits to store the classical data. The total number of qubits in each QNN model is equal to $n_{\text{tot}}=n_{\text{data}}n_{\text{copy}}$. We use $C\equiv n_{\text{gate}}n_{\text{obs}}$ to represent the circuit complexity where $n_{\text{gate}}$ is the number of quantum gates in the variational quantum circuit and $n_{\text{obs}}$ is the number of observables. For QCL and CCQ, $n_{\text{copy}}$, $n_{\text{lay}}$ are free to choose; for DQNN, $n_{\text{lay}}$ and $n_{\text{obs}}$ are free to choose. We choose two types of DQNN in the simulations: DQNN$_1$ contains one circuit and multiple observables, while DQNN$_4$ contains four circuits and one observable. All of the results are obtained by averaging $10$ independent simulations.}
\label{table_regression}
\end{table*}

\begin{figure*}[htb]
    \subfigure[]
    {
     \begin{minipage}{.21\linewidth}\label{qcl_dqnn}
     \centering
      \includegraphics[width=1\textwidth]{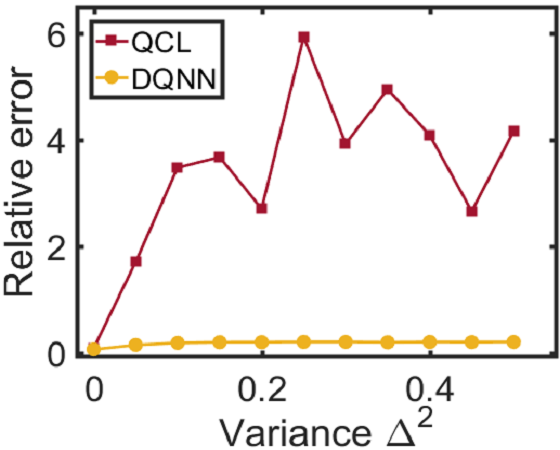}
     \end{minipage}
    }
    \subfigure[]
    {
     \begin{minipage}{.21\linewidth}\label{ccq_dqnn}
     \centering
      \includegraphics[width=1\textwidth]{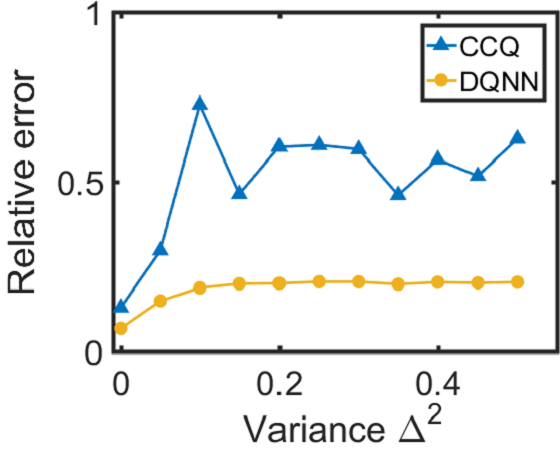}
     \end{minipage}
    } 
    \subfigure[]
    {
     \begin{minipage}{.21\linewidth}\label{circle_noise}
     \centering
      \includegraphics[width=1\textwidth]{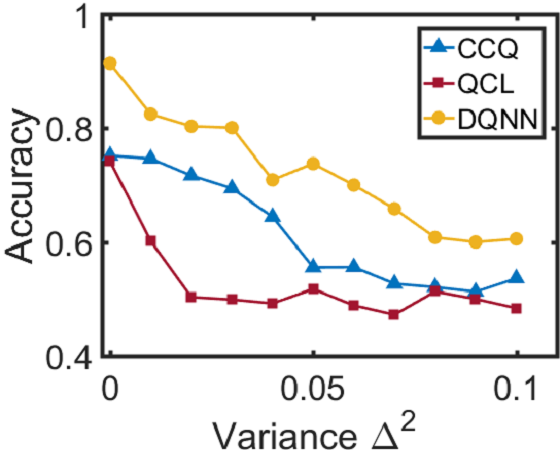}
     \end{minipage}
    }
    \subfigure[]
    {
     \begin{minipage}{.21\linewidth}\label{dephasing_noise}
     \centering
      \includegraphics[width=1\textwidth]{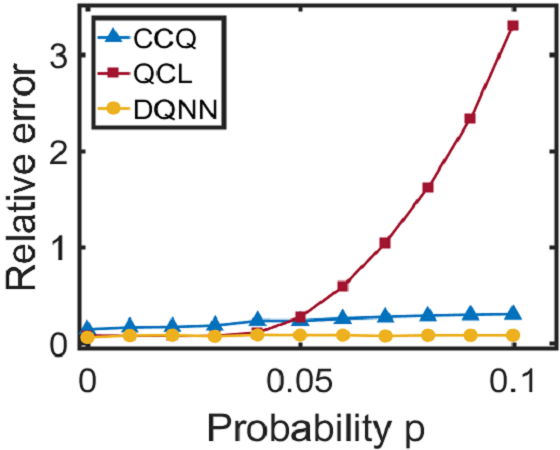}
     \end{minipage}
    }
   \caption{Comparison of the robustness for three QNN models in solving the regression and classification problems. In the regression task, we use DQNN$_1$, $2$-qubit QCL and $2$-qubit CCQ for the comparison. In the classification task, we use DQNN$_1$, $4$-qubit QCL and $4$-qubit CCQ for the comparison.  \textbf{(a)-(b)} Relative errors achieved by QCL, CCQ and DQNN as the coherent noise increases in the regression task. \textbf{(c)} The comparison among three QNNs in the classification task as the coherent noise increases. \textbf{(d)} The comparison among three QNNs in the regression task as the probability of decoherence occurrence increases.}
\end{figure*}

Since the DQNN can reduce the circuit complexity, we expect that it can achieve a better performance in the presence of noise, and hence is easier to be implemented on NISQ devices. We will consider two types of noise in this work: one is the coherent noise caused by the imprecision of the classical control on the parameter values in the QNN circuits; the other is the decoherence generated by interactions between the quantum register and its environment. We add a coherent noise $\epsilon_j$ to the $j$-th parameter, $\theta_j$, in the variational circuit according to $\tilde{\theta}_j=\theta_j+\epsilon_j$ when we train the three QNN models, and test their performance under the noisy environment. The coherent noise $\epsilon_j$ is generated by sampling from a Gaussian distribution, $N(0,\Delta^2)$ where $\Delta$ indicates the noise intensity. It can be seen from Fig.~\ref{qcl_dqnn} and~\ref{ccq_dqnn} that for the regression task, the DQNN achieves lower relative error than QCL and CCQ when we increase the coherent noise. Meanwhile, for the classification task (as shown in Fig.~\ref{circle_noise}), DQNN is more robust than the other two when the coherent noise increases. In addition to the coherent noise, we further apply the decoherence, $\epsilon$, to the two-qubit control gates, $U_{CR}$, contained in the quantum circuits according to $\epsilon(U_{CR}\rho U_{CR}^{\dagger})=(1-p)U_{CR}\rho U_{CR}^{\dagger}+p\sum P_i U_{CR}\rho U_{CR}^{\dagger} P_i^{\dagger}$ where $\{P_i\}$ indicates the complete Pauli basis and $p$ represents the probability of decoherence occurrence. From Fig.~\ref{dephasing_noise}, we can find that, in the regression task, the DQNN has a significantly lower relative error than QCL and CCQ as the probability of decoherence increases. These numerical results demonstrate that the DQNN can decrease the influence of noise accumulation in the training process and is more likely to be implemented on near-term devices.

Besides the comparison among three QNN models over two supervised learning problems, we further demonstrate the performance of the different QNNs in solving the typical real-world machine learning problem based on the handwritten digits data set, MNIST~\cite{lecun2010}. In this task, a QNN is first trained with $60,000$ figures, each of which has a number ranging from $0$ to $9$, and then is tested with $10,000$ new figures to generate the classification accuracy. A figure in the MNIST data set is reshaped to be a $256$-dimensional vector and encoded into an $8$-qubit quantum state by using the amplitude encoding method. Since the QCL needs $256$ qubits to encode the classical data, it is intractable to implement the QCL in solving this task and we only compare the performance of the DQNN and the CCQ. For our DQNN, we use $20$ $2$-layer variational circuits and one observable $B=Z^{(1)}\otimes Z^{(2)}\otimes\cdots \otimes Z^{(8)}$. The classification accuracy over $10,000$ new figures can achieve $94.36\%$. For the CCQ, we use one variational quantum circuit and a set of POVM operators $\{B_j\}_{j=1}^{10}$ to generate the final output. We adopt different layers for the variational quantum circuit, however, the performance of the CCQ is still significantly lower than the DQNN. For instance, the CCQ with a $20$-layer variational quantum circuit only achieves $83.52\%$ accuracy. The result demonstrates that the QNN without a strict universality guarantee will fail in solving machine learning problems. 

\section{Comparison with classical neural networks}
To demonstrate that our DQNN can perform at least as effectively as the classical NN, we apply the DQNN to solve three typical real-world classification problems and further compare its performance with the classical NN in the minimum parameters required the number of iterations to achieve a similar classification accuracy. Since our DQNN is essentially with a single hidden layer, we will choose a single-hidden-layer classical neural network for the comparison. On the handwritten digits data set, MNIST~\cite{lecun2010}, we implement a binary classification~(MNIST$_2$) to distinguish the digits $0$ and $1$, and a $3$-target classification (MNIST$_3$) to recognize the digits $0$, $1$, $2$. Each sample is reshaped into a $256$-dimensional vector and taken as the input of DQNN and classical NN. Meanwhile, we implement two more classification tasks based on the Wine data set~\cite{Dua:2019} and the Breast Cancer data set~\cite{Dua:2019}. We use a $4$-qubit state to store the $13$-dimensional classical data for the Wine problem,  and a $4$-qubit state to store the $10-$dimensional data for the Breast Cancer problem. As shown in Table~\ref{Actual set}, the DQNN performs as well as the classic neural networks in discovering complex patterns hidden in real-world data sets. Notice that DQNN uses fewer parameters to achieve similar accuracy than the classical counterpart but is at the extra cost of required iterations in some cases.

\begin{table*}[htp]
\centering
    \begin{tabular}{c|c|c|c|c|c|c|c}
    \hline
    \hline
    Model & Task & $N_{\text{Train}}$ & $N_{\text{Test}}$ & \# Parameters & \# Iterations & $\text{Acc}_{\text{Train}}$ & $\text{Acc}_{\text{Test}}$ \\
    \hline
    DQNN & MNIST$_2$ &$12665$ & $2115$ & $54$ & $150$ & $98.66\%$ & $99.05\%$\\
    \hline
    Classical NN & MNIST$_2$ &$12665$ & $2115$ & $259$ & $26$ & $98.54\%$ & $99.07\%$\\
    \hline
    DQNN & MNIST$_3$ & $18623$ & $3147$ & $495$ & $254$ & $98.61\%$ & $99.03\%$\\
    \hline
    Classical NN & MNIST$_3$ & $18623$ & $3147$ & $523$ & $367$ &$98.48\%$ & $99.02\%$\\
    \hline
    DQNN & Wine & $143$& $35$ & $27$ & $172$ & $96.29\%$ & $100\%$ \\
    \hline
    Classical NN & Wine & $143$& $35$ & $83$ & $230$ & $97.04\%$ & $100\%$ \\
    \hline
    DQNN & Breast Cancer & $560$& $39$ & $34$ & $282$ &$94.46\%$ & $95.74\%$\\
    \hline
    Classical NN & Breast Cancer & $560$& $39$ & $55$ & $105$ &$93.57\%$ & $95.89\%$\\
    \hline
    \hline
    \end{tabular}
    \caption{A comparison of the minimum parameters required and the number of iterations to achieve a similar classification accuracy between DQNN and classical neural network on three typical real-world data sets. We use the ADAM algorithm to optimize the parameters. $N_{\text{Train}}$ and $N_{\text{Test}}$ respectively denote the number of the training set and the test set. $\text{Acc}_{\text{Train}}$ and $\text{Acc}_{\text{Test}}$ represent the classification accuracy on each data set.}
    \label{Actual set}
\end{table*}

\section{Quantum phase recognition}
\begin{figure}[htp]
    \centering
    \includegraphics[width=6cm,height=4.5cm]{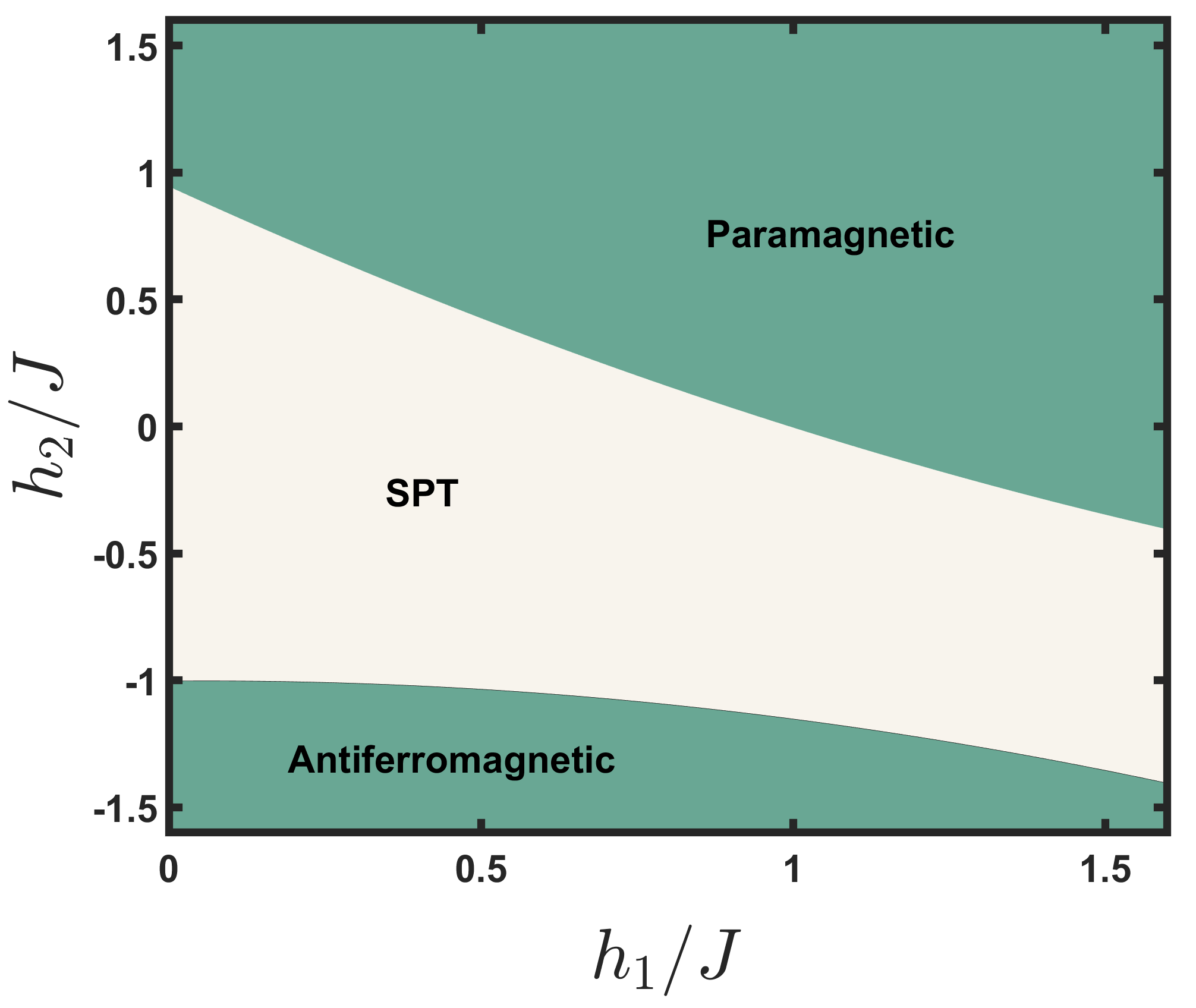}
   \caption{ The phase diagram of the $S=1$ Haldane chain. The phase boundary is generated by the 2-degree polynomial regression based on some boundary points.}
    \label{phase}
\end{figure}

Besides the classical ML problems, our DQNN model can solve ML problems on quantum data sets. In the following, we concentrate on a class of quantum phase recognition problems, the discrimination of the $\mathbb{Z}_2 \times \mathbb{Z}_2$ symmetry-protected topological (SPT) phase on a $S=1$ Haldane chain~\cite{haldane1983nonlinear}. The ground state of the Hamiltonian,
\begin{align*}
    H=&-J\sum_{i=1}^{N-2} Z^{(i)} X^{(i+1)} Z^{(i+2)}-h_1\sum_{i=1}^N X^{(i)}\\
    &-h_2\sum_{i=1}^{N-1} X^{(i)}X^{(i+1)},
\end{align*}
has three topological phases, the SPT phase, the paramagnetic phase and the antiferromagnetic phase, where $h_1$ and $h_2$ are parameters. Our goal is to identify whether a given state sampled from the phase diagram in Fig.~\ref{phase} belongs to the SPT phase. To complete such a task, we take $20\times20$ equally spaced points from $h_1\in[0,1.6]$ and $h_2\in[-1.6,1.6]$ as the training set and $64\times64$ equally spaced points as the test set. If the ground state belongs to the SPT phase, it is labeled $[1,0]^T$; otherwise, it is labeled $[0,1]^T$. We numerically implement the DQNN with a single $15$-qubit variational quantum circuit with $420$ parameters and $10$ observables. After training, the accuracy of the DQNN on the test data achieves $99.10\%$. It indicates that our DQNN model has solved the quantum phase recognition problem with good performance. However, the training set to train the DQNN is larger than the training set to train the quantum convolutional neural network~\cite{cong2019quantum} for phase recognition task. The reason for the difference is that the DQNN used in the comparison is only a prototype, while the quantum convolutional neural network in~\cite{cong2019quantum} has been optimized for this task, with a stronger ability to analyze the local information of quantum states. Hence, as a future work, we hope to improve the DQNN design to optimize its performance for the given ML problem. 

\section{Conclusion}
There have been many successes in achieving larger scale and higher fidelity of the NISQ devices. These improvements provide the quantum neural network with the opportunity to deal with larger data sets and even outperform the classical counterpart. This paper presents a duplication-free quantum neural network model, DQNN, to solve machine learning problems. Meanwhile, we provide it with the universal guarantee to approximate any arbitrary continuous function using several variational quantum circuits, multiple measurement observables and the classical parameterized sigmoid function. We can enhance the expressibility of DQNN by increasing the number of quantum circuits and the number of observables without requiring auxiliary qubits. This property for the number of qubits makes DQNN more likely to be implemented on intermediate-scale quantum computers.

Unlike other variational QNN proposals, DQNN utilizes the classical computer in the hybrid system to generate nonlinearity. Without duplicates, DQNN significantly reduces the number of required qubits and decreases circuit complexity compared with two well-known QNN models. Our simulation results demonstrate that, compared with QCL and CCQ, DQNN uses fewer qubits and fewer quantum gates to outperform the other two QNN models and hence weaken the influence of the circuit noise. Furthermore, DQNN requires fewer parameters than the classical counterpart but is at an extra cost of the iterations in some cases. DQNN can also be used to solve quantum problems. These results indicate that the DQNN is an efficient QNN model which can find the patterns hidden in the classical and quantum data sets.

Like the classical naive neural network, DQNN can be considered a subroutine to construct more complex quantum deep learning models. In addition, DQNN can also be improved by optimizing its structure for the given ML problem. For example, DQNN can be combined with the concept of recurrent neural network to efficiently solve the natural language processing tasks, or integrated into a quantum reinforcement learning framework. Moreover, due to its simple design, DQNN is relatively easy to be implemented on NISQ devices for QNN demonstration.

\section*{Acknowledgements}
The authors gratefully acknowledge the grant from the National Key R\&D Program of China, Grant No. 2018YFA0306703. We also thank Chu Guo, Bujiao Wu, Yusen Wu, Shaojun Wu, Yuhan Huang, Donghong Han, Yingli Yang and Yi Tian for helpful discussions.

\appendix
\begin{figure*}[htp]
    \centering
    \includegraphics[width=0.7\textwidth]{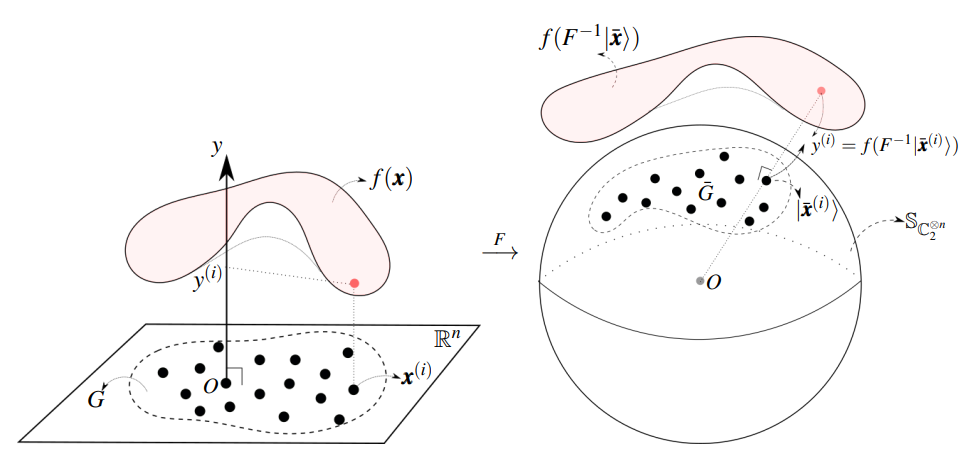}
\caption{Via the amplitude encoding process $F:\bm{x} \mapsto |\bar{\bm{x}}\rangle$, approximating $f(\bm{x}):G \rightarrow \mathbb{R}$ is equivalent to the the problem of approximating $f(F^{-1}|\bar{\bm{x}}\rangle)$,
which can be achieved efficiently by DQNN.}
\label{fig:fx-f-bar}
\end{figure*}
\section{Amplitude encoding method}\label{Appendix_amplitude_encoding}
Suppose $\{\vec{x}^{(i)}\}$ of the data set $D=\{\vec{x}^{(i)},y^{(i)}\}$ is contained in a subset $G$ of $\mathbb{R}^d$. Without loss of generality, we can choose an appropriate coordinate frame through translation such that $\bm 0\notin G$ and every $\boldsymbol{x} \in G$ satisfies $0< \kappa_1 \leq |\boldsymbol{x}| \leq \kappa_2$ with constants $\kappa_1$ and $\kappa_2$.
In order to apply variational QNN, we encode the classical data into the states of a quantum register. In this work, we define the encoding mapping $F : G \rightarrow \bar{G}$ such that $\ket{\bar{\bm x}} = F(\boldsymbol{x})$ is an $n$-qubit state in the Hilbert space $\mathbb{C}_2^{\otimes n}$ with $d\le 2^n$:
\begin{equation}\label{mapping}
    F:\bm{x}=(x_1,...,x_d)^T \mapsto |{\bar{\bm x}}\rangle \equiv \gamma^{-1}(x_1,...,x_d,\tilde{x},0,...,0)^T
\end{equation}
where $\tilde{x}=\frac{\|\boldsymbol{x}\|}{1+\|\boldsymbol{x}\|}$, $\gamma=(\|\boldsymbol{x}\|^2+\tilde{x}^2)^{\frac{1}{2}}$. It is easy to validate that $\frac{\kappa_1}{1+\kappa_1}\leq \tilde{x} \leq \frac{\kappa_2}{1+\kappa_2}$ and
\begin{equation}\label{Eq:x-d+1}
(1+(1+\kappa_2)^2)^{-\frac{1}{2}}<\bar{x}_{d+1}<(1+(1+\kappa_1)^2)^{-\frac{1}{2}} 
\end{equation}
with $\bar{x}_{d+1} = \gamma^{-1} \tilde{x}$. Such encoding method is called amplitude encoding, since the classical information $\bm{x}$ is encoded into the amplitudes of $\ket{\bar{\bm x}}$, as illustrated in Fig.~\ref{fig:fx-f-bar}.
Therefore, approximating $f(\bm{x}): G \rightarrow \mathbb{R}$ is equivalent to the approximation of $f(F^{-1}\ket{\bar{\bm x}}): \bar{G} \rightarrow \mathbb{R}$.

\section{Proof of Theorem 1}\label{Appendix:Proof_theorem1}

It suffices to show $Q(\bar{G})=\{q_{\bm{z}_1,\cdots,\bm{z}_{n_s},\bm{a},\bm{c},\bm{\alpha}} \}$ is dense in $L^2(\bar{G})$. We will prove it by contradiction. Assume that the closure $\overline{Q(\bar{G})}\neq L^2(\bar{G})$. By the Hahn-Banach theorem~\cite{hahn1927lineare}, there exists a bounded linear functional $l$ on $L^2(\bar{G})$ such that $l(Q(\bar{G}))=0$ and $l\neq 0$. By the Riesz representation theorem~\cite{frechet1907ensembles}, there exists a function $g(\bar{\bm{x}})\in L^2(\bar{G})$ such that the linear functional $l$ can be represented as
\begin{equation}
    l(f)=\int_{\bar{G}}f(\bar{\boldsymbol{x}})g(\bar{\boldsymbol{x}}) d\mu \qquad \text{for all } f\in L^2(\bar{G}).
    \label{eq:lf_g}
\end{equation}
Note that $l \neq 0$ implies $g \neq 0$ on $\bar{G}$. 
Therefore, the following two sets $S_1=\{\bar{\boldsymbol{x}} \in \bar{G} : g(\bar{\boldsymbol{x}}) >0  \}$  and $S_2=\{ \bar{\boldsymbol{x}} \in \bar{G} : g(\bar{\boldsymbol{x}}) < 0\}$ are measurable and at least one of them has a positive measure. Without loss of generality, we assume that $g(\bar{\boldsymbol{x}})>0$ almost everywhere in an open subset $E \subset \bar{G}$ with $\mu(E)>0$.
Substituting $f(\bm x)=\sigma(a(|\langle\bar{\bm{x}}|\bm{z}\rangle|^2-c))\in Q(\bar{G})$ into above equation, and in view of $l(Q(\bar{G}))=0$, we have: 
\begin{equation}\label{eq:prf-1}
    \int_{\bar{G}} \sigma(a(|\langle\bar{\boldsymbol{x}}|\boldsymbol{z}\rangle|^2-c))g(\bar{\boldsymbol{x}})d\mu=0.
\end{equation}
As the open set $E$ contains a small ball $B(\boldsymbol{z}^*,\delta)=\{\bar{\boldsymbol{x}} \in \bar{G} : |\bar{\boldsymbol{x}}-\boldsymbol{z}^*|<\delta\}$,  
we will show that, by choosing large enough $a>0$ and $0<c<1$, the function $\sigma(a(|\langle\bar{\boldsymbol{x}}|\boldsymbol{z}\rangle|^2-c))$ tends to $1$ in a smaller ball $B(\boldsymbol{z}^*,\delta_1) \subset B(\boldsymbol{z}^*,\delta)$ $(\delta_1 < \delta)$ and vanishes quickly outside of $B(\boldsymbol{z}^*,\delta_1)$, which leads to a contradiction to Eq.~\eqref{eq:prf-1}.

Since $\bar{G}$ is a real subset of the unit sphere, 
\begin{equation}
    |\bar{\boldsymbol{x}}-\boldsymbol{z}^*|^2 =2-2\langle\bar{\boldsymbol{x}}|\boldsymbol{z}^*\rangle \qquad \text{ for all } \bar{\boldsymbol{x}} \in \bar{G},  
\end{equation}
which implies that $|\langle\bar{\boldsymbol{x}}|\boldsymbol{z}^*\rangle|^2>c$ is equivalent to one of the following two cases:
\begin{equation}\label{Eq:x-xi-x+xi}
       \text{(i)} \ |\bar{\boldsymbol{x}}-\boldsymbol{z}^*|^2<2(1-\sqrt{c}) \quad \text{or} \quad \text{(ii)} \ |\bar{\boldsymbol{x}}-\boldsymbol{z}^*|^2>2(1+\sqrt{c}).
\end{equation}
For $c \to 1$, case (i) means $\bar{\boldsymbol{x}}$ is sufficiently close to $\boldsymbol{z}^*$, and (ii) means $\bar{\boldsymbol{x}}$ is sufficiently close to $-\boldsymbol{z}^*$, but the latter case is impossible. Specifically, for $c$ sufficiently close to $1$, and a given value $\kappa_2>0$, 
we can choose $c$ such that $c \geq (1-2(1+(1+\kappa_2)^2)^{-1})^2$, and together with Eq.~(\ref{Eq:x-d+1}), we have:
\begin{equation}\label{contra}
|\bar{\boldsymbol{x}}+\boldsymbol{z}^*|^2 \geq (z^*_{d+1} + \bar{x}_{d+1})^2\geq 4(1+(1+\kappa_2)^2)^{-1}\ge 2(1-\sqrt{c})
\end{equation}
If (ii) were true, then by $|\bar{\boldsymbol{x}}-\boldsymbol{z}^*|^2+|\bar{\boldsymbol{x}}+\boldsymbol{z}^*|^2=4$, we would find $|\bar{\boldsymbol{x}}+\boldsymbol{z}^*|^2<2(1-\sqrt{c})$, contradictory to Eq.~(\ref{contra}). Hence case (ii) is impossible. Therefore, when $c\rightarrow 1$, $|\langle\bar{\boldsymbol{x}}|\boldsymbol{z}^*\rangle|^2>c$ is only equivalent to case (i). Passing to the limit $a\rightarrow+\infty$, we get 
\begin{equation}\label{eq-sig-limit}
    \sigma(a(|\langle\bar{\boldsymbol{x}}|\boldsymbol{z}^*\rangle|^2-c))\rightarrow \left\{
    \begin{aligned}
        &1 \quad \text{ for $\bar{\vec x}$ with } |\bar{\boldsymbol{x}} - \boldsymbol{z}^*|<\delta_1, \\
        &0 \quad \text{ for $\bar{\vec x}$ with } |\bar{\boldsymbol{x}} - \boldsymbol{z}^*|>\delta_1,
    \end{aligned}
\right.
\end{equation}
with $\delta_1=(2(1-\sqrt{c}))^{\frac{1}{2}}$. Taking $c$ sufficiently close to $1$, we have $\delta_1\leq\delta$, and from Eq.~(\ref{eq:prf-1}) we obtain
\begin{align}
    0&=\int_{\bar{G}} \sigma(a(|\langle\bar{\boldsymbol{x}}|\boldsymbol{z}\rangle|^2-c))g(\bar{\boldsymbol{x}})d\mu \\
    &\rightarrow \int_{B(\boldsymbol{z}^*,\delta_1)}1 g(\bar{\boldsymbol{x}}) d\mu > 0 \quad (\text{as } a\rightarrow +\infty), 
\end{align}

which is a contradiction. Hence, we conclude that $Q(\bar{G})$ is dense in $L^2(\bar{G})$. Thus for any $f \in L^2(\bar{G})$ and $\epsilon>0$, there exists a $q(\bar{\boldsymbol{x}}) \in Q(\bar{G})$ such that
\begin{equation}
    \|f-q\|^2_{L^2(\bar{G})}=\int_{\bar{G}}|f(\bar{\boldsymbol{x}})-q(\bar{\boldsymbol{x}})|^2 d\mu \leq \epsilon, 
\end{equation}
which proves Theorem~\ref{theorem1}. 
\begin{remark}
If $\sigma$ is a ReLU-type function, then the conclusion still holds. 
In fact, instead of Eq.\eqref{eq-sig-limit}, we have $\sigma(a(|\langle\bar{\boldsymbol{x}}|\boldsymbol{z}\rangle|^2-c)) > 0$ for all $ \bar{\boldsymbol{x}}\in B(\boldsymbol{z}^*,\delta_1)$ and $\sigma(a(|\langle\bar{\boldsymbol{x}}|\boldsymbol{z}\rangle|^2-c)) = 0$ otherwise, which leads to the contradiction: 
\begin{align*}
    0&=\int_{\bar{G}} \sigma(a(|\langle\bar{\boldsymbol{x}}|\boldsymbol{z}\rangle|^2-c))g(\bar{\boldsymbol{x}})d\mu \\
    &= \int_{B(\boldsymbol{z}^*,\delta_1)} \sigma(a(|\langle\bar{\boldsymbol{x}}|\boldsymbol{z}\rangle|^2-c))g(\bar{\boldsymbol{x}})d\mu> 0.
\end{align*}
\end{remark}

\section{Introduction to other variational QNN models}\label{Appendix:CCQ_QCL}

In order to compare the performance of DQNN with other models, in this section, we will present a brief introduction to two established variational QNN models, circuit-centric quantum classifier~(CCQ)~\cite{schuld2020circuit} and quantum circuit learning~(QCL)~\cite{mitarai2018quantum}, and will apply these models to solve the following quantum learning problem: constructing a QNN to approximate an $M$-order polynomial $f$: $\vec x\to f(\vec x)$, where $M$ is an even number. 

First, we consider the CCQ model, where the classical data $\bm{x}=[x_1,x_2,\cdots,x_d]^T\in \R^d$ with $||\bm{x}||=1$ is mapped into a $n_{\text{data}}$-qubit register with $n_{\text{data}}\equiv \lceil \log d \rceil$, using the following encoding scheme:
\begin{equation}\label{amplitude_encoding}
    |\bar{\vec{x}}\rangle=\frac{1}{\gamma}[x_1,x_2,\cdots,x_d,\tilde{x},0,\cdots0]^T \in \CC_2^{\otimes n_{\text{data}}},
\end{equation}
where $\tilde{x}$ is a padding term chosen by the user, and $\gamma$ is the normalization factor. Notice that $\ket{\bar{\vec{x}}}$ alone is not sufficient as an input to generate the desired nonlinearity, and an input state as a tensor product of copies of $\ket{\bar{\vec{x}}}$ is required to generate the high-order terms in $f(\vec x)$. Specifically, to approximate the polynomial $f$, the input state $\ket{\psi_{\vec x}}$ of CCQ is chosen as: 
\begin{align}
    |\psi_{\vec x}\rangle&=|\bar{\vec{x}}\rangle^{\otimes n_{\text{copy}}}=|\bar{\vec{x}}\rangle^{\otimes\frac{M}{2}}\\
    &=\frac{1}{\gamma^{\frac{M}{2}}}[x_1^{\frac{M}{2}},x_1^{\frac{M}{2}-1}x_2,\cdots,\tilde{x}^{\frac{M}{2}},0,\cdots,0]^T \in \CC_2^{\otimes n_{\text{tot}}},\label{psi_x_ccq}
\end{align}
where $n_{\text{tot}}\equiv n_{\text{copy}}n_{\text{data}}=\frac{1}{2}M \lceil \log d \rceil$. Here we have chosen the number of copies of $\ket{\bar{\vec{x}}}$ in $\ket{\psi_{\vec x}}$ to be $n_{\text{copy}}=\frac{M}{2}$. 

\begin{figure*}[htp] 
\centering 
\includegraphics[width=0.7\textwidth]{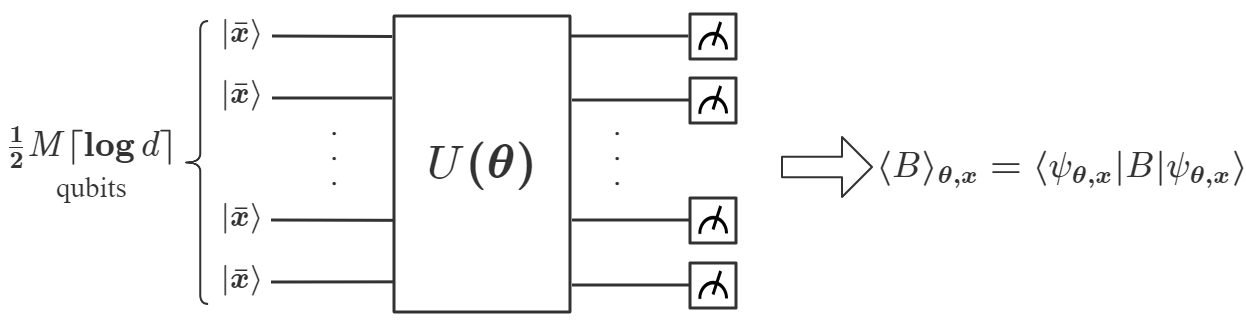} 
\caption{The quantum circuit for CCQ to approximate the $M$-order polynomial function $\vec x\to f(\vec x)$.} 
\label{CCQ_frame} 
\end{figure*}

Denoting the variational quantum circuit in CCQ by $U(\vec\theta)$, the output state $\ket{\psi_{\vec\theta,\vec x}}$ becomes:
\begin{widetext}
\begin{equation}
    \label{CCQ_evolve}
    \ket{\psi_{\vec\theta,\vec x}}=U(\vec\theta) |\psi_{\vec x}\rangle=\frac{1}{\gamma^{\frac{M}{2}}}
    \left[
    \begin{matrix}
    u_{1,1}(\vec\theta) & u_{1,2}(\vec\theta) & u_{1,3}(\vec\theta) & \cdots & u_{1,2^{n_{\text{tot}}}}(\vec\theta)\\
    u_{2,1}(\vec\theta) & u_{2,2}(\vec\theta) & u_{2,3}(\vec\theta) & \cdots & u_{2,2^{n_{\text{tot}}}}(\vec\theta)\\
    u_{3,1}(\vec\theta) & u_{3,2}(\vec\theta) & u_{3,3}(\vec\theta) & \cdots & u_{3,2^{n_{\text{tot}}}}(\vec\theta)\\
    \vdots & \vdots & \vdots & \ddots & \vdots\\
    u_{2^{n_{\text{tot}}},1}(\vec\theta) &  u_{2^{n_{\text{tot}}},2}(\vec\theta) &  u_{2^{n_{\text{tot}}},3}(\vec\theta) &\cdots & u_{2^{n_{\text{tot}}},2^{n_{\text{tot}}}}(\vec\theta)
    \end{matrix}
    \right]
    \left[
    \begin{matrix}
    x_1^{\frac{M}{2}}\\
    x_1^{\frac{M}{2}-1}x_2\\
    x_1^{\frac{M}{2}-1}x_3\\
    \vdots \\
    0
    \end{matrix}
    \right]
    =   
    \left[
    \begin{matrix}
    p_1(\vec\theta,\vec x)\\
    p_2(\vec\theta,\vec x)\\
    p_3(\vec\theta,\vec x)\\
    \vdots \\
    p_{2^{n_{\text{tot}}}}(\vec\theta,\vec x)
    \end{matrix}
    \right]
\end{equation}
\end{widetext}
where each amplitude $p_i(\vec\theta,\vec x)$ of $\ket{\psi_{\vec\theta,\vec x}}$ is an $\frac{M}{2}$-order polynomial of $\vec{x}$. Hence, for a chosen observable $B$, the measurement outcome $\langle B\rangle_{\vec\theta,\vec x}\equiv\bra{\psi_{\vec\theta,\vec x}}B\ket{\psi_{\vec\theta,\vec x}}$ is an $M$-order polynomial of $\vec{x}$. Then the goal of CCQ is to choose an appropriate $B$ and to find the best value $\vec\theta=\vec\theta^*$ through quantum-classical hybrid optimization such that $\vec x\to \langle B\rangle_{\vec\theta^*,\vec x}$ is the optimal function to approximate $\vec x\to f(\vec x)$. The quantum circuit for CCQ is presented in Fig.~\ref{CCQ_frame}.

Next, we consider the QCL model, where the data point $\bm{x}=[x_1,x_2,\cdots,x_d]^T$ is encoded into a $d$-qubit data-encoding pure-state density operator:
\begin{align}
        \rho(\vec{x})&=\ket{\psi_{{\vec x}}}\bra{\psi_{{\vec x}}} \nonumber\\
        &=\otimes_{i=1}^d \big( R_{i}^Y(\arcsin x_i)|0\rangle_i\langle0|_iR_{i}^{Y\dag}(\arcsin x_i) \big) \nonumber\\
        &=\frac{1}{2^d}\otimes_{i=1}^d \big(I+x_i X+\sqrt{1-x_i^2}Z \big)
        \label{QCL_encoding}
\end{align}
where $R_i^Y(\vec\theta)$ represents a Pauli-$Y$ rotation on the $i$-th qubit. Analogous to the CCQ case, $\rho(\vec{x})$ along is not sufficient as an input to generate the desired nonlinearity. In order to generate an $M$-order polynomial, $\rho(\vec{x})$ is duplicated into $M$ identical copies in a tensor product. Specifically, the input state of QCL is chosen as:
\begin{align}
    \rho_{\text{in}}(\vec{x})&=\otimes_{j=1}^M\rho (\vec{x})\nonumber\\
    &=\frac{1}{2^{Md}}\otimes_{j=1}^M\big(\otimes_{i=1}^d [I+x_i X+\sqrt{1-x_i^2} Z\big).
    \label{rho}
\end{align}
Applying a parameterized quantum circuit $U(\vec\theta)$ to $\rho_{\text{in}}(\vec{x})$, followed by a measurement of a chosen observable $B$, we obtain the measurement outcome as the output of QCL: 
\begin{equation}\label{measurement}
    \langle B \rangle_{\vec\theta,\vec x}\equiv\Tr\big(U(\vec\theta)\rho_{\text{in}}(\vec{x})U^{\dagger}(\vec\theta)B\big)
\end{equation}

\begin{figure*}[htp] 
\centering 
\includegraphics[width=0.8\textwidth]{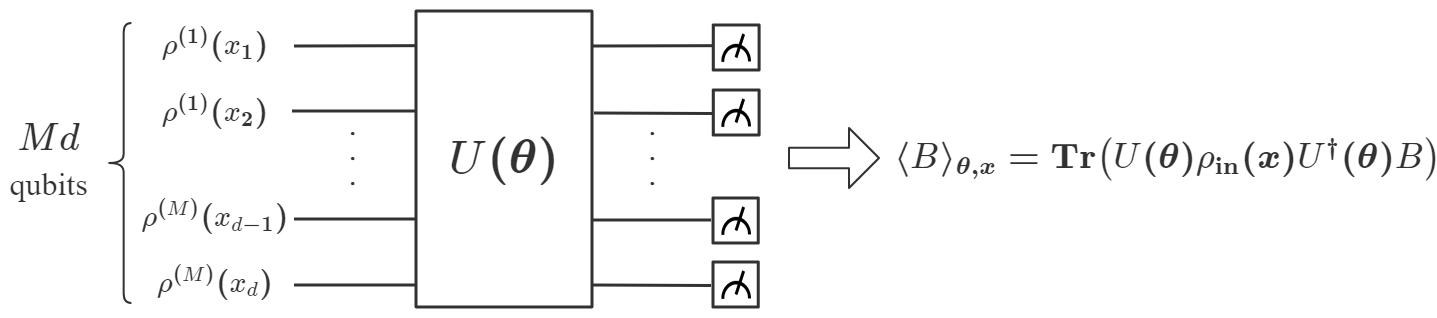} 
\caption{The quantum circuit for QCL to approximate the $M$-order polynomial function $\vec x\to f(\vec x)$.} 
\label{QCL_frame} 
\end{figure*}

Define $\{A_l\}\equiv \{I, X, Z\}^{\otimes Md}$ as a set of generalized Pauli operators. Due to Eq.~(\ref{rho}), $\rho_{\text{in}}(\vec{x})$ can be expanded in terms of $\{A_l\}$:
\begin{equation}\label{inputs}
    \rho_{\text{in}}(\vec{x})=\frac{1}{2^{Md}}\sum_{l=1}^{3^{Md}}p_l(x_1,\cdots,x_d,\sqrt{1-x_1^2},\cdots,\sqrt{1-x_d^2}) A_l,
\end{equation}
where $(x_1,\cdots,x_d,x_1',x_2',\cdots,x_d')\in\RR^{2d} \to p_l(x_1,x_2,\cdots,x_d,x_1',x_2',\cdots,x_d')$, $l=1,2,\cdots,3^{Md}$, are polynomials with respect to all $2d$ variables. Assuming $B$ is unitarily diagonalized as
\begin{equation}\label{observable}
    B=W^{\dagger}\left[\begin{matrix}
        \lambda_1 & \cdots & 0\\
        \vdots    & \ddots & \vdots\\
        0         & \cdots & \lambda_{2^{n_{\text{tot}}}}
    \end{matrix}
    \right]W,
\end{equation}
we substitute Eq.~(\ref{inputs}) and Eq.~(\ref{observable}) into Eq.~(\ref{measurement}) and obtain: 
\begin{align*}
    \langle B \rangle_{\vec\theta,\vec x}&=\frac{1}{2^{Md}}\sum_{l=1}^{3^{Md}}b_l(\vec\theta,\lambda)p_l\\
    b_l(\vec\theta,\lambda)&\equiv\sum_{j=1}^{2^{Md}}\lambda_j\text{Tr}\big(A_l(WU(\vec\theta))^{\dagger}|e_j\rangle\langle e_j|WU(\vec\theta)\big),
\end{align*}
where $\{|e_j\rangle\}$ is the set of computational basis. Thus, the task of QCL is to choose an appropriate observable $B$ and to find the best value $\vec\theta=\vec\theta^*$ through quantum-classical hybrid optimization such that $\vec x\to \langle B\rangle_{\vec\theta^*,\vec x}$ will well approximate $\vec x\to f(\vec x)$. The quantum circuit for QCL is given in Fig.~\ref{QCL_frame}.

In summary, to prepare the desired initial state to solve the same polynomial approximation problem, it requires $\frac{1}{2}M\lceil \log d \rceil$ qubits for CCQ, and $Md$ qubits for QCL. In comparison, it only requires $\lceil \log d \rceil$ qubits for DQNN to prepare the input state as $\ket{\bar{\vec x}}$ in Eq.~(\ref{amplitude_encoding}). This is a significant reduction for large $M$ and large $d$. Specifically, for DQNN, after we prepare the input state $\ket{\bar{\vec x}}$, we sequentially apply a series of variational quantum circuits, $\{U^{(j)}(\vec\theta^{(j)})\}_{j=1}^{n_{\text{cir}}}$, and quantum measurement to derive the measurement outcome $\langle B_i \rangle_{\vec\theta^{(j)},\vec x}=\langle \bar{\vec x}|U^{(j)\dagger}(\vec\theta^{(j)})B_iU^{(j)}(\vec\theta^{(j)})|\bar{\vec x}\rangle$ for a set of chosen observables $\{B_i\}_{i=1}^{n_{\text{obs}}}$. Next, we apply a parameterized classical sigmoid function $\sigma$ to each $\langle B_i \rangle_{\vec\theta^{(j)},\bar{\vec x}}$ to construct the final output $q_{\vec\theta, \vec\alpha,\vec a,\vec c}(\bar{\vec x})$. Finally, we implement the hybrid optimization to find the optimal values of $(\vec\theta,\vec\alpha,\vec a,\vec c)$ to approximate the target polynomial, where $\vec \alpha$, $\vec a$, $\vec c$ are parameters used to construct the final output $q_{\vec\theta, \vec\alpha,\vec a,\vec c}$, with details in the following. 

\section{Solving regression problems using DQNN, CCQ and QCL}\label{Appendix:regression}

In the following, we aim to solve a regression task of approximating a highly nonlinear function:
\begin{align}
    f:(x_1,x_2)\in [-0.8,0.8]^2 \to &(0.716 - 1.0125 x_1^2 + x_1^4)\nonumber\\
    &\times(0.716 - 1.0125 x_2^2 + x_2^4)\in \RR
    \label{regression_prob}
\end{align}   
using DQNN, QCL and CCQ. The data set $D\equiv\{\vec x^{(i)}, y^{(i)}\}=\{(x_1^{(i)},x_2^{(i)},y^{(i)})\}_{i=1}^{400}$ is of size $400$, with $(x_1^{(i)},x_2^{(i)})$ randomly chosen from the square $[-0.8,0.8]^2$, and each label $y^{(i)}$ is calculated according to $y^{(i)}=f(x_1^{(i)},x_2^{(i)})$. The goal of a QNN is to generate a nonlinear function $\vec x \to q_\vec\theta(\vec x)$ to approximate $f$ by optimizing over the parameters $\vec\theta$ of the QNN. 

\begin{figure*}[htp]
    \centering
    \subfigure[DQNN$_1$ based on one single-layer quantum circuit and $4$ observables.]{\includegraphics[width=0.8\textwidth]{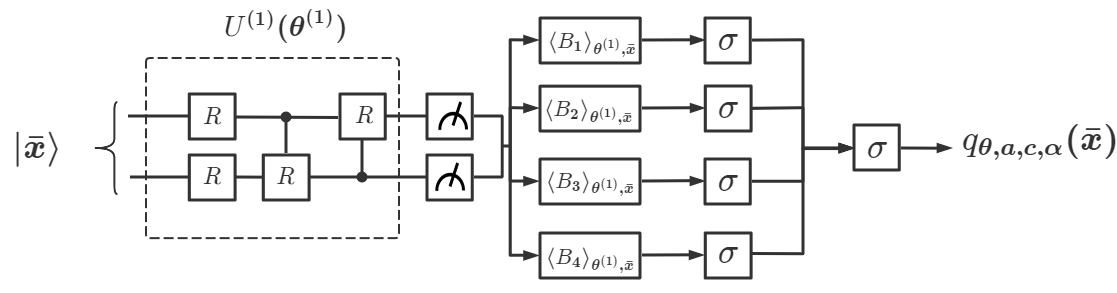}
   }
   \\
    \subfigure[DQNN$_4$ based on $4$ single-layer quantum circuits and one observable.]{\includegraphics[width=0.8\textwidth]{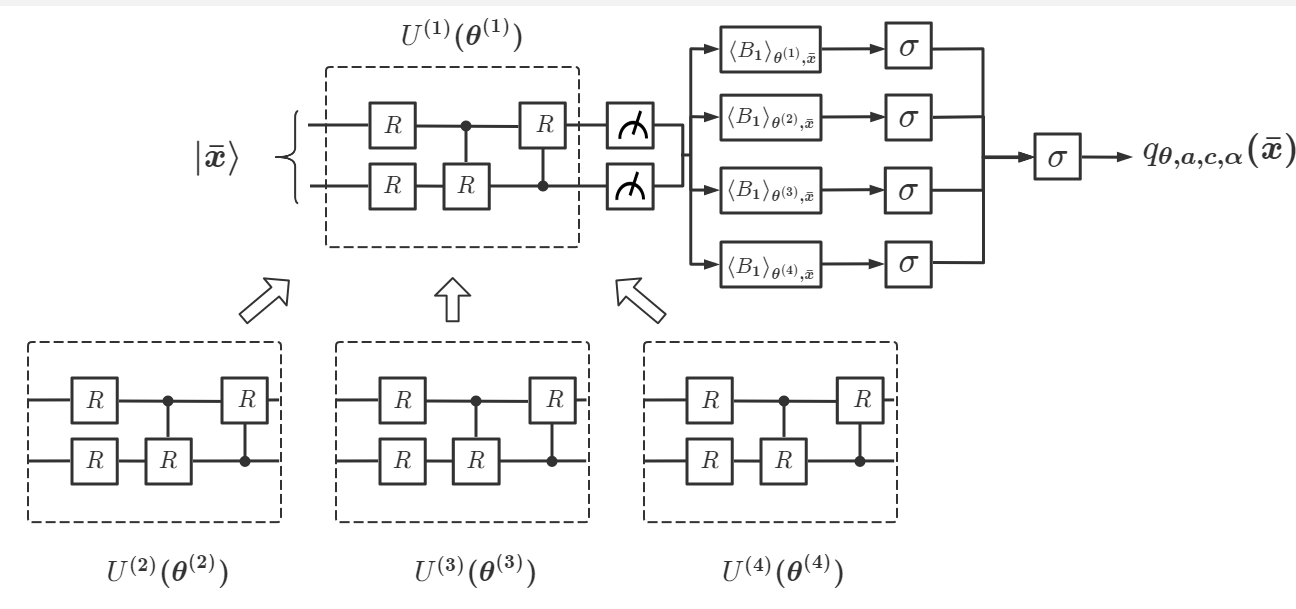}}
    \caption{The implementation of DQNN in the regression task of Eq.~(\ref{regression_prob}). }
    \label{DQNN_REG}
\end{figure*}

First, we use the DQNN model to solve this problem. We choose two cases for DQNN in the task. Case $1$~(DQNN$_1$) is based on a $2$-qubit single-layer variational quantum circuit, $U^{(1)}(\bm{\theta}^{(1)})$, and four measurement observables, $\{B_i\}_{i=1}^4$, randomly chosen the generalized Pauli basis on two qubits, as shown in Fig.~\ref{DQNN_REG}(a). Case $2$ (DQNN$_4$) is based on four $2$-qubit single-layer variational quantum circuits, $\{U^{(j)}(\bm{\theta}^{(j)})\}_{j=1}^4$, and one measurement observable $B_1=Z^{(1)}\otimes Z^{(2)}$, as shown in Fig.~~\ref{DQNN_REG}(b). Through state reparation, each data point $\vec x$ is encoded into the input state:
\begin{equation}\label{amp_encoding}
    |\bar{\vec{x}}\rangle=\frac{1}{\gamma}[x_1,x_2,\tilde{x},0]^{T}
\end{equation}
where $\tilde{x}=\frac{|\boldsymbol{x}|}{1+|\boldsymbol{x}|}$ and $\gamma=(|\boldsymbol{x}|^2+\tilde{x}^2)^{\frac{1}{2}}$. Then we sequentially apply the variational circuit $U^{(j)}(\vec\theta^{(j)})$ and quantum measurement to obtain the measurement outcome $\langle B_i \rangle_{\vec\theta^{(j)},\bar{\vec x}}$:
\begin{align}
\langle B_i\rangle_{\vec\theta^{(j)},\bar{\bm{x}}}\equiv \mathrm{Tr}(U^{(j)}(\vec\theta^{(j)})|\bar{\bm{x}}\rangle\langle \bar{\bm{x}}|{U^{(j)}(\vec\theta^{(j)})}^{\dagger}B_i).
\end{align}
Denoting the sigmoid function as $\sigma(x)$, with
\begin{equation}
    \sigma(x)\equiv \frac{1}{1+e^{-x}},
\end{equation}
for DQNN$_1$, we obtain
\begin{align}
    u(\bar{\bm{x}})\equiv \sum_{i=1}^{4}\alpha_i^{(1)}\sigma(a^{(1)}_i(\langle B_i\rangle_{\vec\theta^{(1)},\bm{\bar{x}}}-c^{(1)}_i)), \, i=1,2,3,4.
\end{align}
and finally derive
\begin{equation}
    q_{\vec\theta,\vec a,\vec c,\vec \alpha}(\bar{\bm{x}})\equiv \sigma(a_{5}(u(\bar{\bm{x}})-c_{5}))
\end{equation}
where the parameter $\vec\theta$ comes from the quantum circuit $U(\vec\theta)$, while the parameters $(\vec a,\vec c,\vec \alpha)$ come from the classical processing after $U(\vec\theta)$.  
For DQNN$_4$, we obtain
\begin{align}
    u(\bar{\bm{x}})\equiv \sum_{j=1}^{4}\alpha_1^{(j)}\sigma(a^{(j)}_1(\langle B_1\rangle_{\vec\theta^{(j)},\bm{\bar{x}}}-c^{(j)}_1)), \, j=1,2,3,4
\end{align}
and derive
\begin{equation}
    q_{\vec\theta,\vec a,\vec c,\vec \alpha}(\bar{\bm{x}})\equiv \sigma(a_{5}(u(\bar{\bm{x}})-c_{5})).
\end{equation}
Finally, through optimization, we can find the optimal values of parameters $(\vec \theta,\vec a,\vec c,\vec \alpha)$ that correspond to the optimal approximation to the target function $f$.  

\begin{figure*}[htp]
    \centering
    \subfigure[][]{\includegraphics[width=0.8\textwidth]{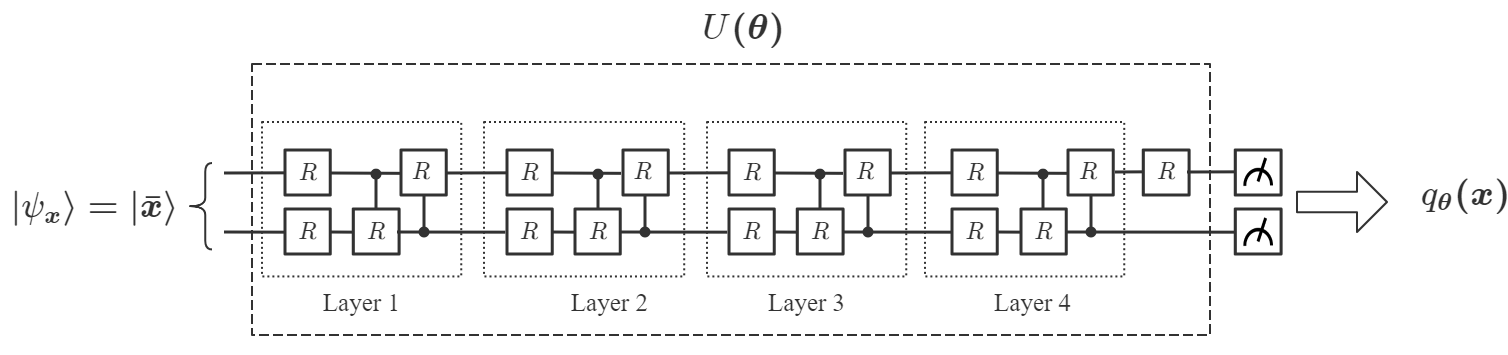}
   }
   \\
    \subfigure[]{\includegraphics[width=0.8\textwidth]{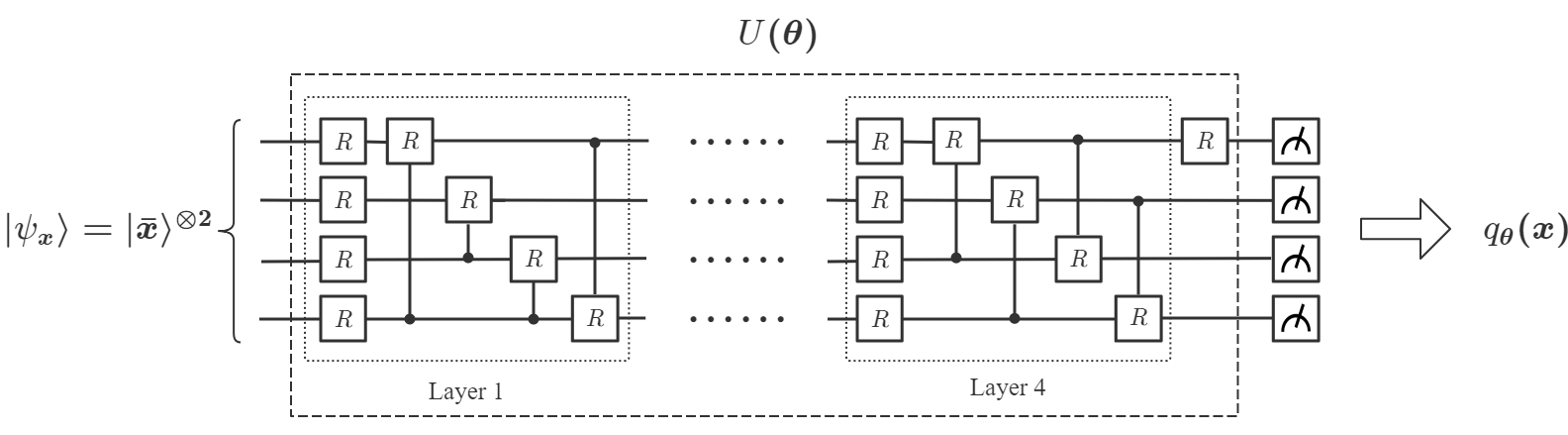}}
    \caption{The circuit of CCQ in the regression task of Eq.~(\ref{regression_prob}), with $U(\vec\theta)$ chosen as a $2$-qubit $4$-layer circuit (case a), and as a $4$-qubit $4$-layer circuit (case b)}
    \label{Fig_CCQ}
\end{figure*}

Second, we consider the CCQ model. For any data point $(\bm{x},y)\in D$, $\bm{x}$ can be encoded into a 2-qubit state $|\bar{\vec{x}}\rangle=\frac{1}{\gamma}[x_1,x_2,\tilde{x},0]^{T}$ as in Eq.~(\ref{amplitude_encoding}). For the input state $\ket{\psi_{\vec x}}$ in Eq.~(\ref{psi_x_ccq}), the number of copies $n_{\text{copy}}$ can be chosen to be $n_{\text{copy}}\ge 1$. Here we consider two cases, $n_{\text{copy}}=1$ and $n_{\text{copy}}=2$. For $n_{\text{copy}}=1$, we design $U(\vec\theta)$ in CCQ to be a $2$-qubit $4$-layer variational quantum circuit; for $n_{\text{copy}}=2$, $U(\vec\theta)$ in CCQ is chosen to be a $4$-qubit $4$-layer circuit. The circuit depth of $U(\vec\theta)$ is $13$ in the former case, and $21$ in the latter case. Correspondingly, the input state $\ket{\psi_{\vec x}}$ is either a 2-qubit state or a 4-qubits state, as shown in Fig.~\ref{Fig_CCQ}. After input state preparation in each case, we apply $U(\vec\theta)$ and a measurement to $\ket{\psi_{\vec x}}$, and derive the measurement outcome
\begin{equation}
    \langle B\rangle_{\vec\theta,\vec x}\equiv \mathrm{Tr}(U(\vec\theta)\ket{\psi_{\vec x}} \bra{\psi_{\vec x}} {U(\vec\theta)}^{\dagger}B)
\end{equation}
where the observable is chosen as $B\equiv Z^{(1)}$, the local Pauli-Z gate on the first qubit. Then we derive the following output 
\begin{equation}
     q_{\vec\theta}(\bm{x})= \frac{\langle B\rangle_{\vec\theta,\vec x}+1}{2}.
\end{equation}
Through optimization, we can find the optimal value  $\vec\theta$ that correspond to the best approximated $q_{\vec\theta}$ to the target function $f$. 

\begin{figure*}[htp]
    \centering
    \subfigure[]{\includegraphics[width=0.8\textwidth]{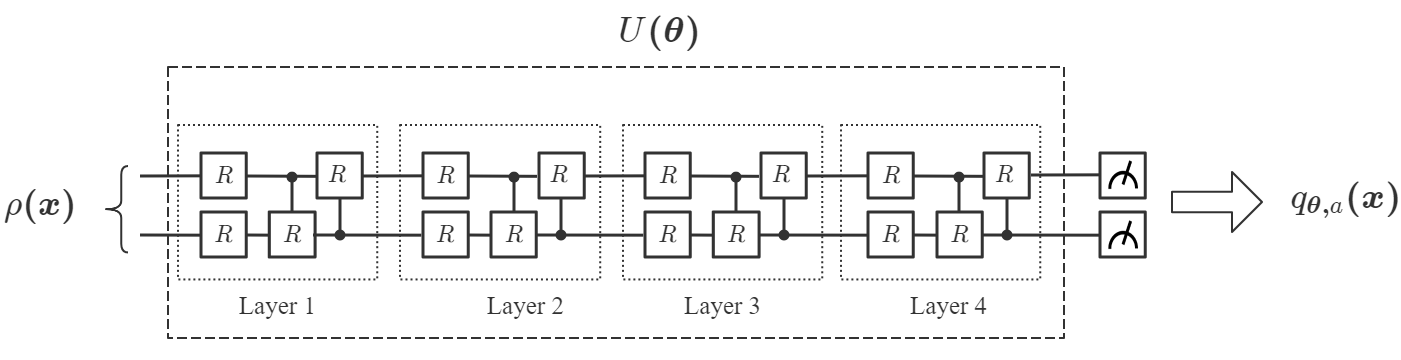}
   \label{QCL_L4C1}}
   \\
    \subfigure[]{\includegraphics[width=0.8\textwidth]{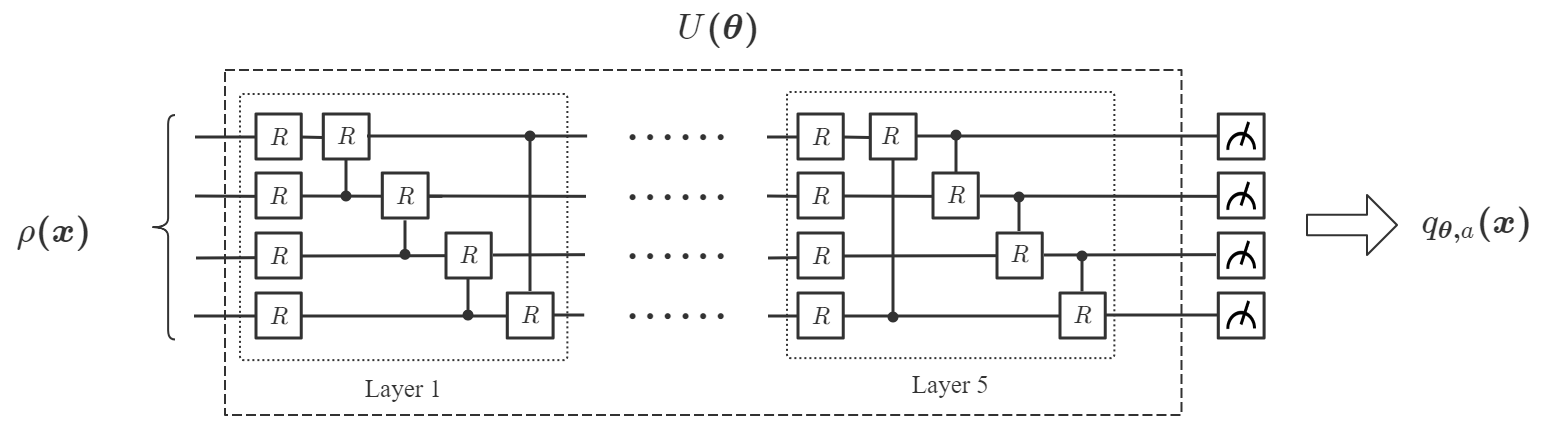}\label{QCL_L5C2}}
    \caption{The circuit of QCL in the regression task of Eq.~(\ref{regression_prob}), with $U(\vec\theta)$ chosen as a $2$-qubit $4$-layer circuit (case a), and as a $4$-qubit $5$-layer circuit (case b).}
    \label{Fig_qcl}
\end{figure*}

Finally, we apply QCL to solve the same problem. We also consider two cases where the classical data point $\vec x$ is encoded into a 2-qubit state ($n_{\text{copy}}=1$) and a 4-qubit state ($n_{\text{copy}}=2$) respectively, as in Eq.~(\ref{QCL_encoding}):
\begin{equation}
     \rho(\vec{x})=\frac{1}{4}\otimes_{j=1}^{n_{\text{copy}}} \otimes_{i=1}^{2}\big(\text{I}+x_i X+\sqrt{1-x_i^{2}} Z\big).
\end{equation}
The entire process from the input state to the final output of QCL is illustrated in Fig.~\ref{Fig_qcl}. Notice that the variational quantum circuit $U(\vec\theta)$ is slightly different from the original suggestion in~\cite{mitarai2018quantum}, as we find that our modified circuit has a better performance. After input state preparation in each case, we apply $U(\vec\theta)$ and a measurement to $\rho(\vec{x})$, and derive the measurement outcome
\begin{equation}
    \langle B\rangle_{\vec\theta,\vec x}\equiv \mathrm{Tr}(U(\vec\theta)\rho(\bm{x}){U(\vec\theta)}^{\dagger}B).
    \label{QCL_measure}
\end{equation}
where the observable is chosen as $B\equiv Z^{(1)}$. Then we obtain the final output of QCL:
\begin{equation}
    q_{\vec\theta,a}(\bm{x})= a \langle B\rangle_{\vec\theta,\vec x}
\end{equation}
where $a$ is a trainable parameter. Then, through optimization, we find the optimal values of the parameters $(\vec\theta,a)$ that correspond to the best approximated $q_{\vec\theta,a}$ to the target function $f$.

\newpage
\bibliography{sample}

\end{document}